\documentclass[aps,pra,twocolumn,showpacs,superscriptaddress,10pt]{revtex4-1}

\usepackage{graphicx}
\usepackage{amsmath}
\usepackage{amsfonts}
\usepackage{amssymb}

\usepackage{graphicx,color}
\usepackage{bm}

\def\bra#1{\langle #1 |}
\def\ket#1{| #1 \rangle}
\renewcommand{\Re}{\mathop{\text{Re}}\nolimits}
\renewcommand{\Im}{\mathop{\text{Im}}\nolimits}
\begin{document}

\title{Quantum Zeno dynamics of a field in a cavity}

\author{J.M.~Raimond}
\affiliation{Laboratoire Kastler Brossel, CNRS, ENS, UPMC-Paris 6, 24 rue Lhomond, 75231 Paris, France}\label{I1}
\author{P.~Facchi}
\affiliation{Dipartimento di Matematica and MECENAS, Universit\`a di Bari, I-70125  Bari, Italy}\label{I1}
\affiliation{INFN, Sezione di Bari, I-70126 Bari, Italy}\label{I1}
\author{B.~Peaudecerf}
\affiliation{Laboratoire Kastler Brossel, CNRS, ENS, UPMC-Paris 6, 24 rue Lhomond, 75231 Paris, France}\label{I1}
\author{S.~Pascazio} 
\affiliation{INFN, Sezione di Bari, I-70126 Bari, Italy}\label{I1}
\affiliation{Dipartimento di Fisica and MECENAS, Universit\`a di Bari, I-70126  Bari, Italy}\label{I1}
\author{C.~Sayrin}
\affiliation{Laboratoire Kastler Brossel, CNRS, ENS, UPMC-Paris 6, 24 rue Lhomond, 75231 Paris, France}\label{I1}
\author{I.~Dotsenko}
\affiliation{Laboratoire Kastler Brossel, CNRS, ENS, UPMC-Paris 6, 24 rue Lhomond, 75231 Paris, France}\label{I1}
\author{S.~Gleyzes}
\affiliation{Laboratoire Kastler Brossel, CNRS, ENS, UPMC-Paris 6, 24 rue Lhomond, 75231 Paris, France}\label{I1}
\author{M.~Brune}
\affiliation{Laboratoire Kastler Brossel, CNRS, ENS, UPMC-Paris 6, 24 rue Lhomond, 75231 Paris, France}\label{I1}
\author{S.~Haroche}
\affiliation{Laboratoire Kastler Brossel, CNRS, ENS, UPMC-Paris 6, 24 rue Lhomond, 75231 Paris, France}\label{I1}
\affiliation{Coll\`{e}ge de France, 11 place Marcelin Berthelot, 75231 Paris, France}\label{I1}

\date{\today{}}

\begin{abstract}
We analyze the quantum Zeno dynamics that takes place when a field stored in a cavity undergoes frequent interactions with  atoms. We show that repeated measurements or unitary operations performed on the atoms probing the field state confine the evolution to tailored subspaces of the total Hilbert space. This confinement leads to non-trivial field evolutions and to the generation of interesting non-classical states, including mesoscopic field state superpositions. We elucidate the main features of the quantum Zeno mechanism in the context of a state-of-the-art cavity quantum electrodynamics experiment. A plethora of effects is investigated, from state manipulations by phase space tweezers to nearly arbitrary state synthesis. We analyze in details the practical implementation of this dynamics and assess its robustness by numerical simulations including realistic experimental imperfections. We comment on the various perspectives opened by this proposal.

\end{abstract}


\maketitle

\section{Introduction}
\label{sec:intro}

The evolution of a quantum mechanical system can be significantly slowed down by a series of frequent measurements~\cite{Zenorev}. 
This effect, named after the Eleatic philosopher Zeno~\cite{Zenoproposal}, has attracted widespread attention during the last 20 years, since Cook proposed to test it on oscillating (two-level) systems~\cite{Cook88}. This was a simplified version of the seminal idea by Misra and Sudarshan~\cite{Zenoproposal}, who had in mind genuinely unstable systems, but it had the important quality of making the quantum Zeno `paradox' (as it was originally considered) amenable to experimental test. 

The quantum Zeno effect (QZE) has been successfully demonstrated in many experiments on various physical systems, such as r.f.\ transitions between ionic hyperfine levels (the first test of Cook's proposal)~\cite{Itano90}, rotation of photon polarization
\cite{kwiat}, Landau-Zener tunneling~\cite{Wilkinson97},
nuclear spin isomers~\cite{Chapovsky}, level dynamics of individual ions
\cite{Toschek}, optical pumping~\cite{molhave2000}, 
preservation of spin polarization in gases~\cite{spinpolarization},
quantum computing qubits undergoing decoherence~\cite{decQC},
Bose-Einstein condensates~\cite{ketterle,becsbargill}, optical systems~\cite{hosten}, NMR~\cite{jones}, 
control of decay in optical waveguides~\cite{optwaveguide}
and cavity quantum electrodynamics (CQED)~\cite{Bernu08}. 
Other experiments have also been proposed, involving neutron spin in a waveguide
\cite{RauchVESTA} and superconducting qubits~\cite{Semba}.

Remarkable applications of the QZE have been realized or proposed, such as the control of decoherence~\cite{Facchi04,decQC}, state purification~\cite{Nakazato04}, implementation of quantum gates~\cite{Shao09} and entanglement protection~\cite{maniscalco}. QZE can also inhibit entanglement between subsystems, making a quantum evolution semi-classical~\cite{Rossi09}. Other proposed applications consist in radiation absorption reduction and dosage reduction in neutron tomography~\cite{neutron_tomography}, control of polarization~\cite{polarization_control}, and other general strategies to control decoherence~\cite{decoherence_control_strategies}.

In all these experiments or experimental proposals, repeated projective measurements block the evolution of the quantum system in a non-degenerate eigenstate of the measured observable, so that the system is frozen by QZE in its initial state. However, more general phenomena can take place, for example when the measurement does not confine the system in a single state, but rather in a multidimensional (quantum Zeno) subspace of its Hilbert space. 
This gives rise to a quantum Zeno dynamics (QZD)~\cite{Facchi02}: the system evolves in the projected subspace under the action of its (projected) Hamiltonian.

No experiment has been performed so far to test the QZD. This would be important in view of possible applications, for example in decoherence and quantum control. We proposed in~\cite{QZDcav} a possible implementation of QZD in a CQED experiment. In this proposal, the field in the cavity undergoes a QZD under the joint action of a coherent source coupled to the mode (responsible for the Hamiltonian coherent evolution) and of a repeated photon-number selective measurement or unitary evolution. This process is based on the spectroscopic interrogation of the dressed levels of a single atom coupled to the cavity mode. These repeated operations create two orthogonal subspaces in the field's Hilbert space, with photon numbers larger or smaller than a chosen value $s$. QZD takes place in one of these subspaces.

We also proposed in~\cite{QZDcav} that this procedure could lead to interesting methods towards the synthesis and manipulation of non-classical states. In this Article, we explore these ideas even further and detail some subtle mechanisms involved in the dynamical evolution inside the Zeno subspace. We show, in particular, that the quantum Zeno dynamics can be used to produce mesoscopic field state superpositions (MFSS), quantum superpositions of coherent components with different  amplitudes. Such highly non-classical states are quite interesting for explorations of the quantum-classical boundary~\cite{Exploring06}.

We start by introducing notations and by sketching the main ideas in Sec.\ \ref{sec:principles}.
We explore the mechanisms of the confined dynamics and introduce the key idea of the `exclusion circle' in phase space in Sec.\ \ref{sec:confined}.
The notion of phase space tweezers and scenarii of state manipulation are analyzed in Sec.\ \ref{sec:tweezers}. Finally, we look at interesting perspectives on state synthesis
in Sec.\ \ref{sec:synthesis}.
We further discuss practical implementation, that can be realized with a state-of-the-art apparatus in Sec.\ \ref{sec:simulations}, where we also compare orders of magnitude and perform a few realistic simulations.
Conclusion and perspectives are given in Sec.\ \ref{sec:concl}.

\section{General principles}
\label{sec:principles}

In this Section, we describe the principle of a quantum Zeno dynamics experiment in the cavity quantum electrodynamics context. 
The first Subsection (\ref{sec:generalities}) exposes the general principle of the method and introduces useful notations.  The method could be used in a variety of experimental settings, and particularly in circuit QED~\cite{circuitQED}. However, for the sake of definiteness, we will discuss it in the framework of a microwave CQED experiment in construction at Ecole Normale sup\'erieure (ENS) involving circular Rydberg atoms and superconducting millimeter-wave cavities. We  discuss the general features of this experiment in Subsection \ref{sec:experimentgeneral}. We then describe how QZD may be implemented in this framework using repeated photon-number selective measurements (\ref{sec:QZDrepmeasurements}) or photon-number selective unitary kicks (\ref{sec:QZDunitary}). 

\subsection{Generalities and notation}
\label{sec:generalities}

A QZD can be achieved either by repeated (possibly unread) measurements of an
observable with degenerate eigenvalues, leading to a non-unitary evolution, or by repeated actions of
a Hamiltonian kick with multidimensional eigenspaces, always leading to a global unitary evolution.
The two procedures can be shown to be equivalent in the $N \to \infty$ limit, where $N$ is the number of operations in a finite time interval $t$~\cite{Facchi04}. For $N$ finite, differences can appear between the unitary and non-unitary procedures. Both measurements and kicks are supposed to take place `instantaneously', namely on a timescale that is the shortest one in the problem at hand. We will discuss here both procedures before focusing on the latter, whose implementation in CQED turns out to be the easiest.

The first procedure consists in $N$ repetitions of a sequence involving the evolution under the action of a Hamiltonian $H$ for a time $\tau=t/N$, generating the unitary $U(\tau)=\exp(-iH \tau/\hbar)$, followed by a projective measurement. The action of this measurement is represented by the projectors $P_\mu$~\cite{schwinger}, corresponding to the obtained result $\mu$ ($\sum_\mu P_\mu=\openone$). If the initial state is contained in the eigenspace associated to $\mu_0$, the measurement gives almost certainly $\mu_0$ for each sequence, in the large $N$ and short $\tau$ limit. The evolution is then confined in the Zeno subspace defined by $P_{\mu_0}$, which is in general multidimensional. The evolution in this subspace reads:
\begin{equation}
\label{evolNmeas}
U_{P}^{(N)}(t)= \left[ P_{\mu_0} U(t/N) \right]^N \to
e^{-iH_{\mathrm{Z}} t /\hbar } P_{\mu_0}\ , 
\end{equation}
for $N\to\infty$, where 
\begin{equation}
\label{HzP}
H_{\mathrm{Z}} = P_{\mu_0} HP_{\mu_0}
\end{equation}
is the Zeno Hamiltonian.

In the second procedure, the system undergoes a stroboscopic evolution, alternating short unitary
evolution steps, governed by $U(\tau)$, with instantaneous unitary `kicks' $U_K$. The succession of $N$ steps yields the unitary:
\begin{equation}
\label{UZN}
U_{K}^{(N)}(t)=[U_KU(t/N)]^N   
\sim U_K^N  e^{-iH_{\mathrm{Z}} t /\hbar }
\end{equation}
for $N\to\infty$, where
\begin{equation}
\label{HZinf}
H_{\mathrm{Z}}=\sum_\mu P_\mu H P_\mu\ ,
\end{equation}
the $P_\mu$s being the (multidimensional) eigenprojections of $U_K$ ($U_KP_\mu=e^{i\lambda_\mu} P_\mu$)~\cite{Facchi02}. 

We observe that, by suitably choosing $P_{\mu_0}$ or $U_K$ in Eqs.\ (\ref{evolNmeas}) and (\ref{UZN}) respectively, one can modify the system evolution by tailoring the QZD, leading to possible remarkable applications.
We shall analyze here both schemes and discuss the experimental feasibility of the procedure (\ref{UZN})-(\ref{HZinf}), related to the so-called `bang-bang' control~\cite{BangBang} used in NMR manipulation techniques~\cite{NMR}. The related mathematical framework is familiar in the context of quantum chaos~\cite{qchaos}.

\subsection{A Cavity-QED setup}
\label{sec:experimentgeneral}

Our proposal for QZD implementation~\cite{QZDcav} is based on the photon-number selective spectroscopic interrogation of the dressed levels for a single atom coupled to a high-quality cavity. In the ENS experiments, a very high-$Q$ superconducting millimeter-wave cavity is strongly coupled to long-lived circular Rydberg states. The long lifetimes of both systems are ideal for the realization of experiments on fundamental quantum effects~\cite{Exploring06}. 

In all experiments realized so far, the atoms were crossing the centimeter-sized cavity mode at thermal velocities ($\simeq$ 250~m/s). The atom-cavity interaction time is thus in the few tens of $\mu$s range. It is long enough to result in an atom-cavity entanglement and short enough so that atoms crossing successively the cavity carry a large flux of information about the field state. This information can be used for the implementation of ideal quantum measurements~\cite{qnd} or for quantum feedback experiments~\cite{feedback}. However, this short interaction time is not compatible with a photon-number selective interrogation of the dressed level structure at the heart of our QZD proposal. 

\begin{figure}
\includegraphics[width=7.5cm]{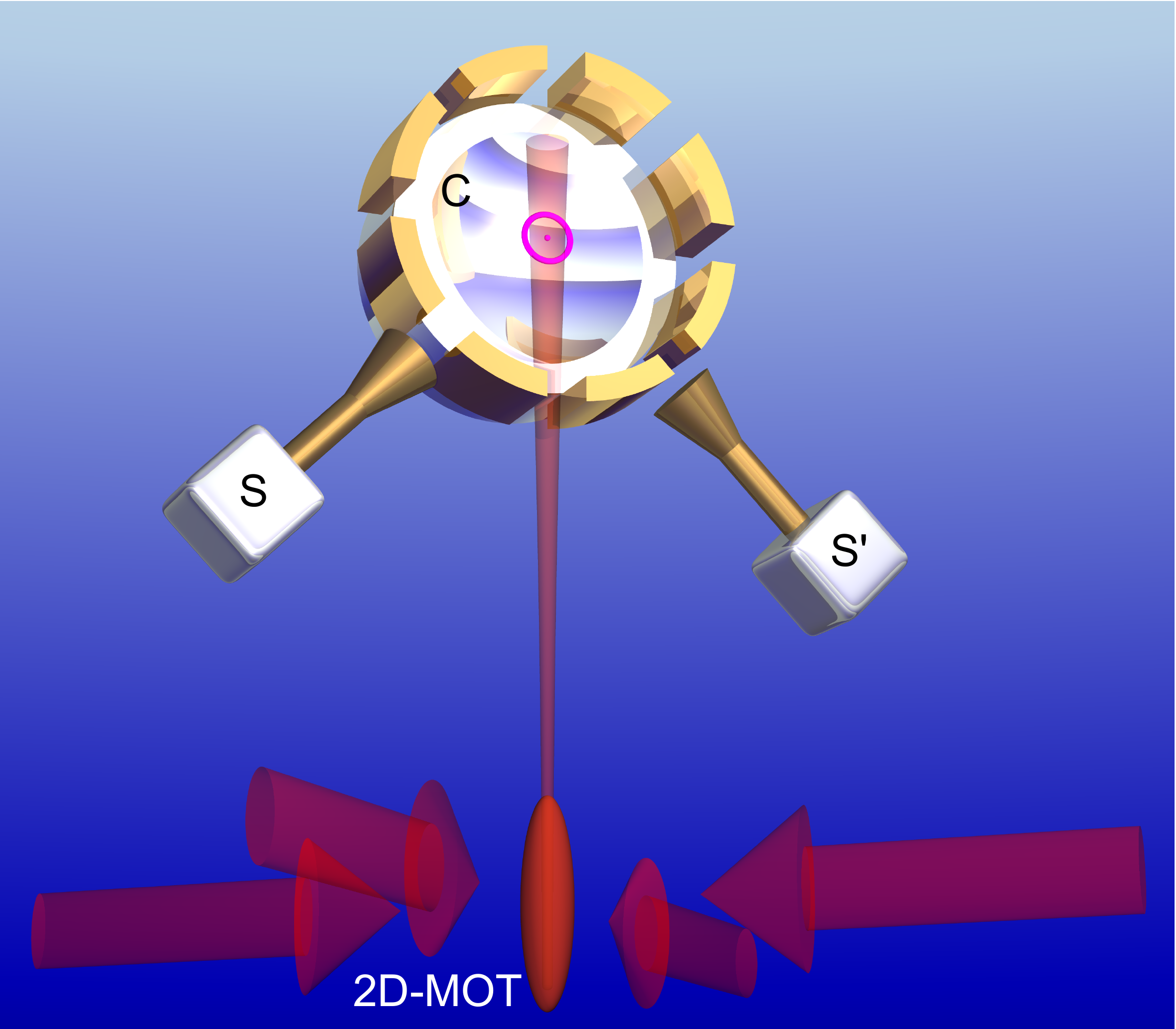}
  \caption{Scheme of the planned ENS cavity QED setup.}
  \label{scheme}
\end{figure}

The ENS group is thus developing a new experiment with slow Rydberg atoms interacting for a long time with the cavity mode. Its scheme is represented on figure~\ref{scheme}. The Fabry-Perot cavity $C$~\cite{cavitytechnique} is made up of two superconducting mirrors facing each other (only one is shown in figure~\ref{scheme} for the sake of clarity). It sustains a non-degenerate Gaussian mode at a frequency close to 51.1~GHz (6~mm wavelength). The mode has a Gaussian standing-wave envelope, with a waist $w=6$~mm. Field energy damping times $T_c$ up to 130~ms have been reached by cooling the mirrors down to 0.8~K. At this temperature, the residual blackbody field corresponds to $n_{th}=0.05$ photons in the mode on the average.

The cavity is resonant with the transition between the two circular Rydberg levels $e$ and $g$, with principal quantum numbers 51 and 50 respectively. These levels have a lifetime of the order of 30~ms, much longer than the typical atom-cavity interaction times considered in this Article (up to a few ms). Atomic relaxation thus plays a negligible role. 

The atoms are prepared by laser and radio-frequency excitation~\cite{Exploring06} out of a slow vertical atomic beam crossing the cavity in an atomic fountain arrangement. A Raman velocity selection performed on the slow beam emanating from a 2D-MOT source placed under the cavity makes it possible to selectively address atoms that are near the turning point of their ballistic trajectory at the cavity center. They thus reside in the mode's waist for a time of the order of 10~ms, limited only by their free fall. 

Excitation lasers are focused in $C$, delimiting a small volume. The initial position of the atoms is thus well-known. The time required for the atomic preparation is short (about 50~$\mu$s). It is important to note that this preparation does not involve any field close to resonance with the cavity mode. It can thus be performed without affecting the field quantum state. 

At the end of their interaction with the field, the atoms can be detected by the field ionization method inside the cavity itself. They are ionized by a field applied across eight electrodes circling the cavity and the resulting ions are routed towards a detector, which produces a macroscopic signal. The method is state-selective, since the ionizing field depends upon the principal quantum number. A simpler scheme can be used to perform an unread detection of the atoms, by merely ionizing them with a field applied directly across the cavity mirrors. Note that the centimeter-sized gaps between the ionizing electrodes enable us to couple millimeter-wave sources to the atoms or to the cavity mode (through its residual diffraction loss channels).

In the following Subsections, we show how this basic setup can be used for implementing the two QZD modes introduced in Section~\ref{sec:generalities}, involving either repeated photon-number selective measurements or repeated photon-number selective fast unitary evolutions.

\subsection{QZD by repeated measurements}
\label{sec:QZDrepmeasurements}

\begin{figure}
\includegraphics[width=4cm]{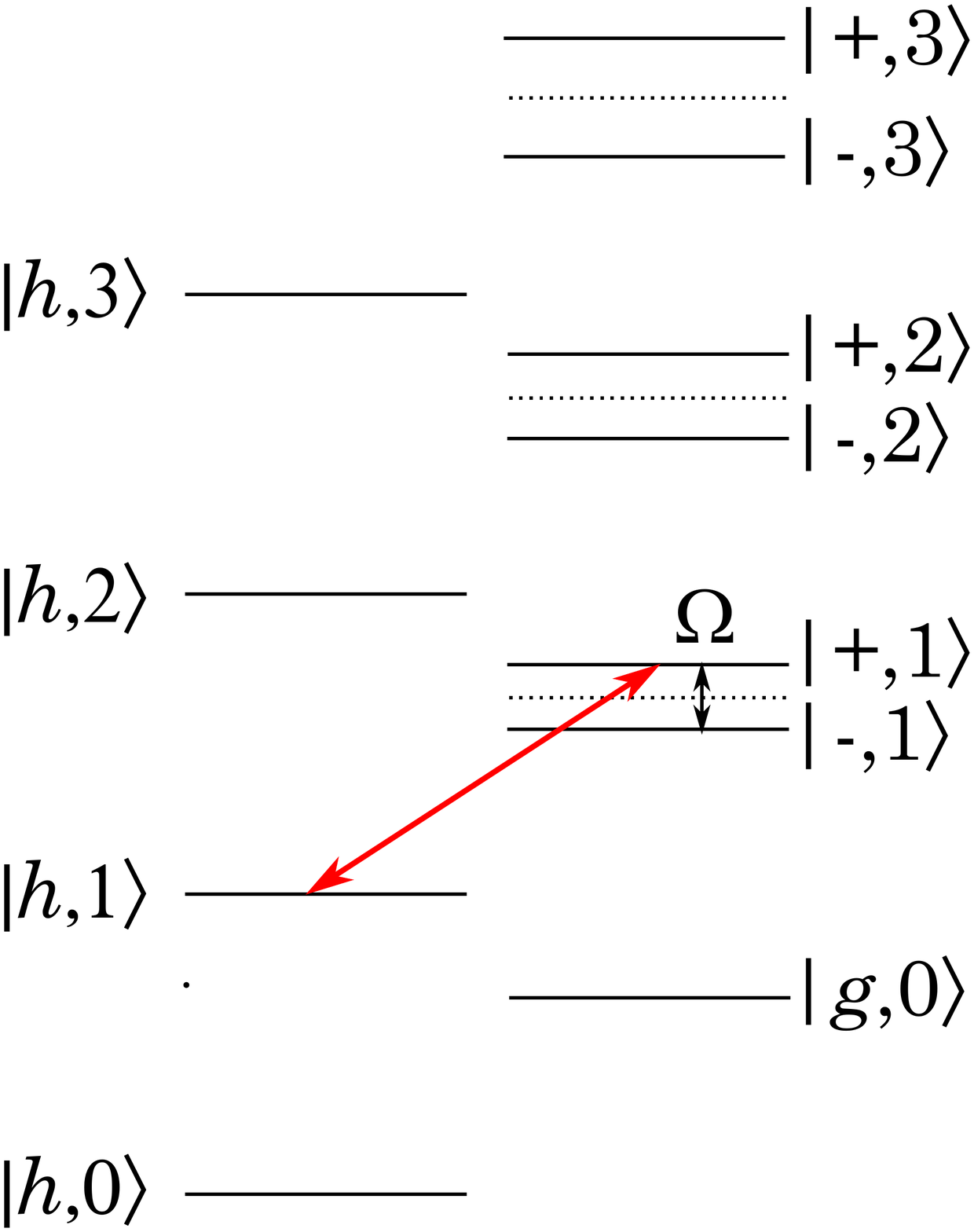}
  \caption{Dressed states of the atom-cavity system. The
arrow indicates the photon-number selective transition addressed
by $S'$ for $s=1$.}
  \label{levelscheme}
\end{figure}

The coherent evolution of the field in $C$ is produced by a classical source $S$  resonantly coupled with  $C$~\cite{Exploring06} (Fig.\ \ref{scheme}). This evolution is
described by the Hamiltonian (we use an interaction representation eliminating the field phase rotation at cavity frequency):
\begin{equation}
\label{Hfree}
H = \alpha a^\dagger + \alpha^* a\ ,
\end{equation}
where $\alpha$ is the source amplitude and $a$ ($a^\dagger$) the photon annihilation (creation) operator. 
If this evolution proceeds undisturbed for a time interval $t$, a coherent state with an amplitude $\xi$ will `accumulate' in the cavity, under the action of the 
unitary displacement operator:
\begin{equation}
\label{UD}
U(t)=D(\xi)=\exp(\xi a^\dagger-\xi^* a)\ , \quad \xi= -i \alpha t/\hbar.
\end{equation}

We now periodically interrupt this evolution over a total time $t$ by $N$ measurements performed at very short time intervals $\tau=t/N$ such that $|\beta|=|-i\alpha\tau/\hbar| \ll 1$.  Each measurement involves a new atom prepared initially in the circular level $h$, with principal quantum number 49. Microwave pulses produced by the source $S'$ probe the transition from $h$ to $g$, at a frequency close to 54.3~GHz. Note that this transition is widely out of resonance from $C$. It can thus be probed without altering the mode state. Moreover, level $h$ is impervious to the cavity field. 

Level $g$ instead is strongly coupled to the cavity mode. In the resonant case, the atom-cavity Hamiltonian in the interaction picture reads:
\begin{equation}
\label{Homega}
V = \frac{\hbar \Omega}{2} ( |e \rangle \langle g| a + |g \rangle \langle e| a^\dagger)\ ,
\end{equation}
where $\Omega$ is the vacuum Rabi frequency ($\Omega/2\pi=50$~kHz). The atom-cavity Hamiltonian eigenstates are the dressed states:
\begin{equation}
\label{Homegaeigen}
\ket{g,0}, \quad
|\pm,n \rangle = \frac{1}{\sqrt{2}} ( |e,n-1 \rangle  \pm |g,n \rangle )\ , \quad n\geq1\ ,
\end{equation}
where the former (latter) entry in each ket refers to the atom (cavity mode).
The splitting between the dressed states $|\pm,n\rangle$ is $\hbar\Omega\sqrt{n}$. 

The pulse sent by the source $S'$ thus actually probes the transition between the level $|h,n\rangle$ (whose energy is independent of the atom-cavity coupling) and the dressed states $|\pm,n \rangle$. The  level structure is shown in Fig.\ \ref{levelscheme}. The frequency of the $|h,n\rangle\rightarrow|+,n\rangle$ transition depends upon the photon number $n$. 

Let us chose a specific photon number $s\ge 1$. The source $S'$ is tuned to perform a $\phi = \pi$ Rabi pulse on the
$|h,s\rangle\rightarrow|+,s\rangle$ transition. It is detuned from the bare $h\rightarrow g$ transition frequency by $\Omega\sqrt{s}/2$. In principle, we can chose the amplitude and duration $\delta t$ of this interrogation pulse so that it has no appreciable effect on the transition between $|h,s\rangle$ and $|-,s\rangle$ or between $|h,n\rangle$ and $|+,n\rangle$ with $n\not=s$. This requires $1/\delta t \ll \Omega|\sqrt{s\pm 1}-\sqrt{s}|$. A long enough atom-cavity interaction time is thus essential for the selective addressing of a single dressed atom transition. 

Finally, the source $S'$ ideally performs the transformations:
\begin{eqnarray}
\label{kick}
U_{s}|h,s \rangle &=& -i |+,s \rangle, \nonumber \\
U_{s}|+,s \rangle &=& -i |h,s \rangle
\end{eqnarray}
and \begin{eqnarray}
\label{kick2}
U_{s}|-,s \rangle &=& |-,s \rangle.
\end{eqnarray}
If the cavity contains a number of photons $n$ different from $s$, then:
\begin{eqnarray}
\label{kickn}
U_{s}|h,n \rangle &=& |h,n \rangle, \nonumber \\
U_{s}|\pm,n \rangle &=& |\pm,n \rangle, \qquad (n \neq s)\ .
\end{eqnarray}
In conclusion,
\begin{eqnarray}
\label{Ukick}
U_{s} &=& -i \left(|h,s \rangle \langle +,s|  + |+,s \rangle \langle h,s| \right) + P_\perp , 
\nonumber\\
P_\perp &=& \openone -  |h,s \rangle \langle h,s| -  |+,s \rangle \langle +,s|\ . 
\end{eqnarray}
This is a unitary process: $U_{s} U^\dagger_{s} = U^\dagger_{s} U_{s}= \openone$\ .

We now examine the global evolution. Assume that the cavity is initially in its ground state and the atom in $h$. The joint atom-cavity state is $|h,0 \rangle$.
After the first time interval $\tau$, it becomes:
\begin{equation}
\label{evolshort}
e^{-i H \tau /\hbar}  |h,0 \rangle =  \sum_{n=0}^\infty c_n(\tau)  |h,n \rangle\ ,
 \end{equation}
where $H$ is given by Eq.\ (\ref{Hfree}). Note that $H$ does not involve any atomic operator. Therefore, $h$ is not affected by the coherent cavity evolution. The coefficients $c_n(\tau)$ are those of a coherent state with a small amplitude $\beta=-i\alpha \tau/\hbar$. After this `free' evolution for a short time $\tau$, the atom undergoes the $\pi$ Rabi pulse driven by $S'$ (\ref{Ukick}):
\begin{equation}
\label{evolshortkick}
U_{s}e^{-iH\tau/\hbar}  |h,0 \rangle =  \sum_{n\neq s} c_n(\tau)  |h,n \rangle - ic_{s}(\tau)  |+,s \rangle\ .
 \end{equation}
At this point, the atom is detected inside the cavity and its state recorded. Since $|\beta|$ is very small, the probability for having $s$ photons or more is small. With a large probability, the atom is thus found in $h$. This is our \emph{measurement}: it makes sure that the number of photons in the cavity is \emph{not} $s$. The cavity field is accordingly almost always projected onto:
\begin{equation}
\label{evolshortkickmeas}
\langle h |U_{s}e^{-iH\tau/\hbar}  |h\rangle =  \sum_{n\neq s} c_n(\tau)  |n \rangle  \ ,
\end{equation}
within a trivial normalization factor. One can thus summarize the action of $U_{s}$ followed by the measurement of $\ket{h}$ by the projection:
\begin{equation}
\label{proj}
P = \openone - |s \rangle \langle s |\ 
\end{equation} 
acting on the field state alone, since
\begin{equation}
P e^{-iH\tau/\hbar}  |0 \rangle = \sum_{n\neq s} c_n(\tau)  |n \rangle\  
\end{equation}
[in terms of operators, we get from (\ref{Ukick}) that, in the photon Hilbert space, $\bra{h} U_{s} \ket{h} = \openone - \ket{s}\bra{s}$].

The Zeno procedure %~(\ref{evolNmeas}) 
consists in the alternating evolution under the action of the free Hamiltonian (\ref{Hfree}) and the projection (\ref{proj}): 
\begin{equation}
\label{evolNkicks}
U_P^{(N)}(t)= \left( P e^{-iH \tau/\hbar } \right)^N, \qquad \tau=t/N\ ,
\end{equation}
which has to be understood as an evolution of the cavity field only.
When $N$ is large,
one gets  $U_P^{(N)}(t)\to U_{\mathrm{Z}}(t)$,
where
\begin{equation}
U_{\mathrm{Z}} = e^{-iH_{\mathrm{Z}} t /\hbar} P\ 
\end{equation}
is the QZD generated by the
the Zeno  Hamiltonian (\ref{HzP}).
Note that 
\begin{equation}
P= P_{<s} + P_{>s}\ , 
\end{equation} 
where $P_{<s} (P_{>s})$
is the projection onto the photon number states with less (more) than $s$ photons. Since $H$ can create or annihilate only one photon at a time, one has $P_{<s} H P_{>s}= 0$, whence 
\begin{equation}
\label{Hz2}
H_{\textrm{Z}} =  P_{<s}HP_{<s} + P_{>s}HP_{>s} =  H_{<s} + H_{>s}\ .
\end{equation}
Here $H_{<s}$ is the restriction of the Hamiltonian $H$ to the photon subspace
$\mathcal{H}_{<s} = P_{<s} \mathcal{H}$, spanned by
the photon number states $|0\rangle,\ldots,|s-1\rangle$, and $H_{>s}$ the restriction to
the subspace $\mathcal{H}_{>s}$ containing more than $s$ photons.

Under the QZD, field states restricted to ${\cal H}_{<s}$ and  ${\cal H}_{>s}$
remain confined in these subspaces, $|s\rangle$ realizing a hard
`wall' between them. Strictly speaking, $P\mathcal{H}= {\cal H}_{<s}  \oplus {\cal H}_{>s}$ forms a single Zeno subspace, within which evolution is coherent. For example, the coherence of the state $(|n\rangle+|p\rangle)/\sqrt{2}$, with $n<s$ and $p>s$ is fully preserved under the Zeno dynamics. However, transitions between ${\cal H}_{<s}$ and ${\cal H}_{>s}$ are forbidden, due to the form of the interaction Hamiltonian. If the initial state is contained in only one of the two sectors, ${\cal H}_{<s}$ or ${\cal H}_{>s}$, it will be confined to it. In the following, we shall focus on this situation. Note that, if $C$ is initially in the vacuum state, with $s=1$, the
system remains inside ${\cal H}_{<1}$, i.e.\ in $|0\rangle$, and we recover the QZE~\cite{Bernu08}.

Of course, for a finite $\tau$ (hence, a finite $\beta$) there is at each step a small but finite probability for finding $s$ photons in the cavity and, hence, the atom in another state than $h$. In this case, the  cavity field is  projected onto the $s$ photon Fock state, and the Zeno procedure is abruptly interrupted. The probability of occurrence of such an event goes to zero when $\tau$ is close to zero. For practical purposes, $\tau$ should be chosen small enough to make this event unlikely in the planned duration of the experiment.

Note also that, as long as the QZD is not interrupted by such an event, the atom is always found in $h$. It is thus not mandatory to actually record the atomic state at the detection time. Merely tracing over the final atomic state leads to the same results for the QZD of the field (in practical terms this means that the atom can be simply ionized by a field applied across the cavity mirrors, before being replaced by a new one for the next QZD step). In fact, as shown in the next Subsection, it is not even necessary to detect the atom at all.

\subsection{QZD by repeated unitary kicks}
\label{sec:QZDunitary}

The same Zeno dynamics can be implemented by making use of a single atom, without any detection. Now, 
$S'$ is tuned to perform a $2\pi$ Rabi pulse on the
$|h,s\rangle\rightarrow|+,s\rangle$ transition. As before, the pulse
amplitude is weak enough (and its duration correspondingly
long enough) not to appreciably affect $|h,n\rangle$ with
$n\not=s$.
This yields the transformation:
\begin{eqnarray}
\label{kick2pi}
U_{s}|h,n \rangle = (-1)^{\delta_{ns}}|h,n\rangle\ ,
\end{eqnarray}
and $U_{s}= \openone$ on all the other states.
The atom always ends up in $h$, while the field experiences the selective kick $U_K=U_s$ with:
\begin{eqnarray}
\label{Ukick2pi}
U_{s} &=& \openone-2|s\rangle\langle s|\ , \\
& & \quad U_{s} U^\dagger_{s} = U^\dagger_{s} U_{s}= \openone\ . \nonumber
\end{eqnarray}
Such a photon-number dependent Rabi
pulse~\cite{Franca87} was used with $s=1$ for a single-photon QND
detection~\cite{QND99} and for a CNOT gate in CQED~\cite{Gate99}.
The evolution (\ref{UZN}) reads:
\begin{equation}
\label{UZN2pi}
U_{K}^{(N)}(t)=[U_sU(\tau)]^N \sim  U_s^N e^{-iH_{\mathrm{Z}} t /\hbar}\ , 
\end{equation}
where $t=N \tau$. The Zeno Hamiltonian (\ref{HZinf}) is:
\begin{eqnarray}
\label{HZinf2pi}
H_{\mathrm{Z}}&=&\sum_{\mu=\pm} P_\mu H P_\mu\ , \\
& & \quad P_-=|s\rangle\langle s|, \quad P_+=P_{<s}+P_{>s}\ ,
\label{HZinf2pibis}
\end{eqnarray}
which also satisfies Eq.~\eqref{Hz2}. Once again, there is a hard
wall at $n=s$, preventing transitions between ${\cal H}_{<s}$ and  ${\cal H}_{>s}$.

\subsection{Interrogation by a generic Rabi pulse}
\label{sec:genphi}

We have seen that a QZD can be obtained both when 
$S'$ drives a $\pi$ Rabi pulse, as in Sec.\ \ref{sec:QZDrepmeasurements}, or a 
$2\pi$ Rabi pulse, as in Sec.\ \ref{sec:QZDunitary}. 
We show here that a generic pulse with an arbitrary Rabi angle $\phi$ yields essentially the same physical effects.

For a generic Rabi pulse, $S'$ performs 
a unitary kick acting on the atom-cavity system, which mixes
$|h,s\rangle$ with $|+,s\rangle$ and would create atom-field
entanglement if $C$ would contain $s$ photons. 
The corresponding unitary operator reads:
\begin{equation}
U_s = \exp\left[-i  \frac{\phi}{2} \big(\ket{h,s}\bra{+,s} + \ket{+,s}\bra{h,s} \big)  \right]\ .
\label{eq:units}
\end{equation}
For $\phi=\pi$, it reduces to the unitary of Sec.~\ref{sec:QZDrepmeasurements}, while it reduces to that of Sec.~\ref{sec:QZDunitary} for $\phi=2\pi$.
The diagonalization of $U_s$ leads to:
\begin{equation}
U_s = e^{-i \frac{\phi}{2}} P_+ + e^{i \frac{\phi}{2}} P_- +  P_\perp\ ,
\end{equation}
with 
\begin{equation}
P_{\pm} = \ket{u_\pm}\bra{u_\pm}\ , \quad
\ket{u_{\pm}} = \frac{\ket{h,s}\pm \ket{+,s}}{\sqrt{2}}\ ,
\end{equation}
and $P_\perp = \openone - P_+ - P_-$.
In the large $N$ limit, for $\phi\neq 0$, the Zeno dynamics is generated by:
\begin{equation}
H_{\mathrm{Z}} = P_+ H P_+ + P_- H P_- + P_\perp H P_\perp\ .
\end{equation} 
Since $\bra{u_\pm} a \ket{u_\pm} =0$, the Zeno Hamiltonian reduces to
\begin{equation}
H_{\mathrm{Z}} = P_\perp H P_\perp\ .
\end{equation}

The unitary~(\ref{eq:units}) admits an
invariant subspace of the range of the eigenprojection $P_\perp$ belonging to the eigenvalue +1. Its
projection is $|h\rangle\langle h|\otimes (P_{<s}+ P_{>s})$, the same
as for a $2\pi$ pulse. Starting from an atom in $|h\rangle$ and a
field in ${\cal H}_{<s}$ or ${\cal H}_{>s}$, we obtain a QZD
leaving the atom in $|h\rangle$ and the field in its initial
subspace. 

Under perfect QZD, the cavity \emph{never} contains $s$
photons and the atom and field are \emph{never} entangled by the
interrogation pulse. This discussion holds in principle for all non-zero values of $\phi$ ($0<\phi\le 2\pi$) in the $N\rightarrow\infty$ limit. For finite values of $N$, QZD  is not properly achieved if $\phi$ is very small, each kick operation being too close to $\openone$. Numerical simulations, to be presented in Section~\ref{sec:limits}, fully confirm this qualitative argument.

\section{Confined dynamics in QZD}
\label{sec:confined}

We have shown that the QZD establishes a hard wall in the Hilbert space, corresponding to the Fock state $|s\rangle$. In qualitative terms, this hard wall can be viewed in the phase space (Fresnel plane) as an `exclusion circle' (EC) with a radius $\sqrt{s}$. In this Section, we examine the QZD starting with an initial coherent field located either inside or outside the EC. This deceptively simple situation leads to the generation of a non-classical MFSS, quantum superposition of distinguishable mesoscopic states.

\subsection{Phase space picture}
\label{sec:phasespace}

\begin{figure*}
\includegraphics[width=17cm]{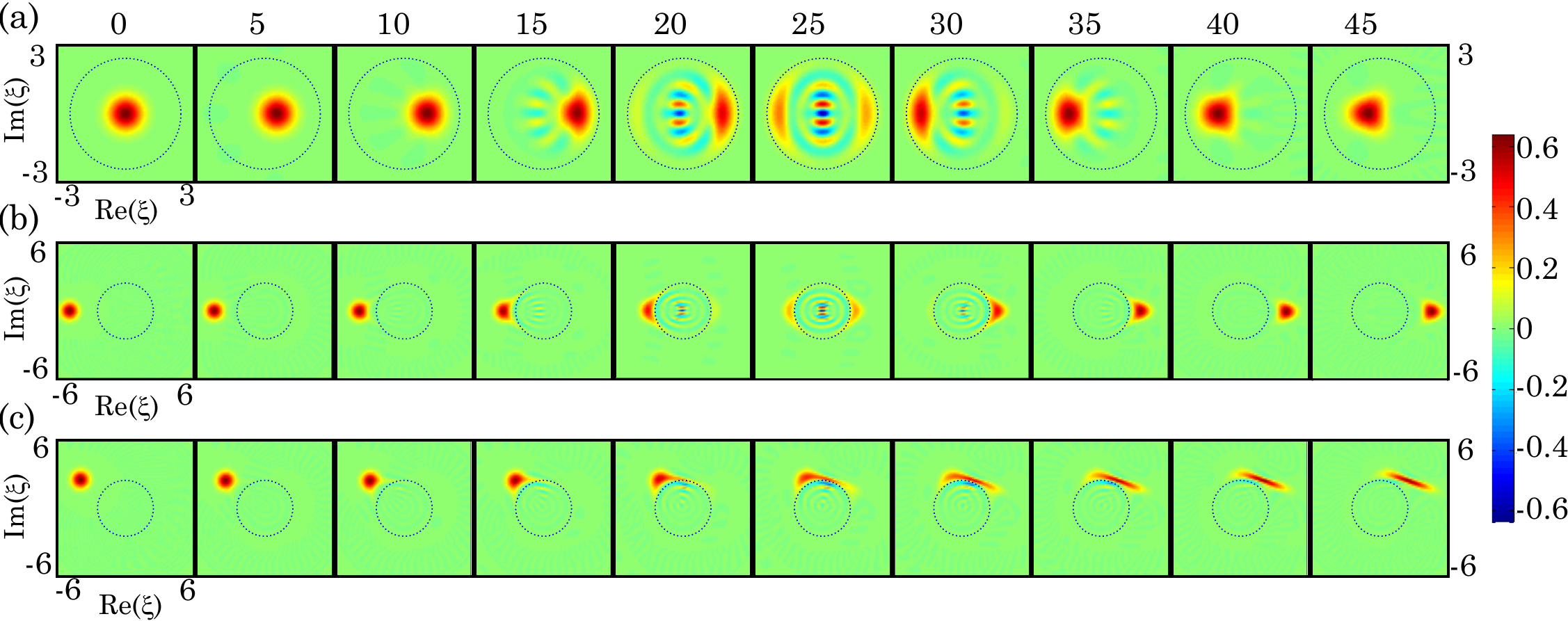}
\caption{(a) QZD dynamics in ${\cal H}_{<6}$. Ten snapshots of the
field Wigner function $W(\xi)$ obtained after a number of steps
indicated above each frame. The cavity is initially in its vacuum
state, $s=6$ and $\beta=0.1$. The EC is plotted as a dashed
line. (b) QZD dynamics in ${\cal H}_{>6}$. Same as (a) with an
initial $\alpha=-5$ amplitude. (c) Same as (b), with an initial
amplitude $\alpha=-4+i\sqrt{6}$. In (b) and (c) the successive
frames correspond to the same step numbers as in (a). From~\cite{QZDcav}.}
  \label{confine}
\end{figure*}

Summarizing the main results of the preceding section, the Zeno dynamics consists in replacing the `free' Hamiltonian (\ref{Hfree}) 
with the Zeno Hamiltonian (\ref{HZinf}):
\begin{equation}
H_{\mathrm{Z}}=\sum_\mu P_\mu H P_\mu  = P_{>s} H P_{>s} + P_{<s} H P_{<s}\ .
\end{equation}
For definiteness, let us assume that $s=4$, identifying the two subspaces 
$\mathcal{H}_Z=P_{<s} \mathcal{H}=\textrm{span}\{|0\rangle, |1\rangle, |2\rangle, |3\rangle\}$
and  ${\cal H}'_\textrm{Z}= \textrm{span}\{|5\rangle, |6\rangle, |7\rangle, \ldots \}$. In the photon-number states basis $\{\ket{n}\}$, the only non-vanishing matrix elements of the annihilation and creation operators are:
\begin{equation}
\bra{n-1}a\ket{n}=\bra{n}a^\dagger\ket{n-1}=\sqrt{n}\ ,
\end{equation}
with $n\geq 1$. The matrix representations of $H$ and $H_{\mathrm{Z}}$ thus read:
\begin{widetext}
\begin{eqnarray}
\label{Hmatrix}
H =\begin{pmatrix}
0&\alpha& 0 & 0 &  0 & 0 & 0  & \hdots      \\
\alpha^*&0& \sqrt{2}\alpha & 0  &  0 & 0 & 0  \\
0 &\sqrt{2}\alpha^*&0 & \sqrt{3}\alpha & 0 & 0 & 0  \\
0 & 0 &\sqrt{3}\alpha^*& 0 &   \sqrt{4}\alpha & 0 & 0  \\
0 & 0 & 0 &  \sqrt{4}\alpha^* & 0 & \sqrt{5}\alpha & 0  \\
0 & 0 & 0 & 0 &  \sqrt{5}\alpha^* & 0 & \sqrt{6}\alpha  \\
0 & 0 & 0 & 0 & 0 &  \sqrt{6}\alpha^* & 0  \\
\vdots &  &  &  &  &   &  & \ddots
\end{pmatrix} 
\stackrel{\textrm{Zeno}}{\longrightarrow}
\begin{pmatrix}
0&\alpha& 0 & 0 &  0 & 0 & 0 & \hdots      \\
\alpha^*&0& \sqrt{2}\alpha & 0  &  0 & 0 & 0  \\
0 &\sqrt{2}\alpha^*&0 & \sqrt{3}\alpha & 0 & 0 & 0  \\
0 & 0 &\sqrt{3}\alpha^*& 0 &  0 & 0 & 0  \\
0 & 0 & 0 &  0 & 0 & 0 & 0  \\
0 & 0 & 0 & 0 &  0 & 0 & \sqrt{6}\alpha  \\
0 & 0 & 0 & 0 & 0 &  \sqrt{6}\alpha^* & 0  \\
\vdots &  &  &  &  &   &  & \ddots
\end{pmatrix} =  H_{\textrm{Z}} , \nonumber \\
\end{eqnarray}
\end{widetext}
respectively.
The two subspaces ${\cal H}_\textrm{Z}$  
and ${\cal H}'_\textrm{Z}$ do not communicate anymore in an ideal QZD situation. If the system is initially in ${\cal H}_\textrm{Z}$, it cannot make a transition to ${\cal H}'_\textrm{Z}$ and \emph{vice versa}. Of course, this simple picture holds only in the limit of a true QZD, with an infinitesimally  small time $\tau$ between kicks and a very small displacement per step $\beta$. We numerically examine the validity of this approximation in Section \ref{sec:limits}.

We simulated in~\cite{QZDcav} the QZD in cavity QED by applying Eq.\ (\ref{UZN2pi}) [as already explained, Eq.\ (\ref{evolNkicks}) would have been equivalent] with the unitary (\ref{UD}) and by making use of Wigner's representation in phase space~\cite{tan}.

We summarize here the main results of our simulations. Figure
\ref{confine} presents three sequences of 10 snapshots of the Wigner
function $W(\xi)$ separated by intervals of 5 steps, for $s=6$
and $\beta=0.1$. The  Wigner function is defined by:
\begin{equation}
W(\xi) = \frac{1}{\pi\hbar} \int \bra{\Re\xi - y} \varrho \ket{\Re\xi + y} e^{i 2 y \Im \xi /\hbar} dy\ ,
\end{equation}
$\varrho$ being the density matrix of the photon field, obtained by tracing out the atomic variable.

In Fig. \ref{confine}(a) the field is initially in its vacuum state $|0\rangle \in {\cal
H}_{<6}$. Its amplitude increases along the real axis (free
dynamics). When this amplitude reaches $\simeq 2$, between 15 and 20 steps, the coherent state `collides' with the EC of radius $\sqrt{6}$ (dashed line in Fig.\ \ref{confine}). The field amplitude stops growing and undergoes a very fast $\pi$ phase shift between steps 20 and 30. At step 25, the field is in  a MFSS, quantum superposition of two
components with opposite phases. The fringes inside the
EC are the signature of the quantum coherence. At step 35, the field state is nearly coherent again with an
amplitude close to $-2$. It then resumes its motion from left to
right along the real axis, going through $|0\rangle$ around
step 45 and heading towards its next collision with the EC. The long-term dynamics will be discussed in Section~\ref{sec:revivals}.

QZD in ${\cal H}_{>6}$ is illustrated in Fig. \ref{confine}(b), with
snapshots of the field Wigner function for $s=6$ and an initial
coherent state $|\alpha=-5\rangle$. The field collides with the EC
after 20 steps. It undergoes a QZD-induced $\pi$ phase shift
being, after 25 steps, in a MFSS. After 30 steps, the state
is again nearly coherent with a positive amplitude and resumes its
motion along the real axis. After 45 steps, its amplitude is
slightly larger than 4.5. It would be $-0.5$ in the case of free
dynamics. 

Finally, in Fig.~\ref{confine}(c), the field state collides
tangentially with the EC. The parts of the Wigner function that come
closest to the EC propagate faster than the others. The state is
distorted and squeezed (albeit by a moderate amount) along one direction.

\subsection{Phase inversion mechanism}

We now show that the main feature of the evolution of the Wigner function, the fast phase shift during the collision with the EC, can be understood via a  semi-classical argument.  
Let us set $\alpha=i/\sqrt{2}$ in (\ref{Hfree}) for simplicity. We get that the Hamiltonian,
\begin{equation}
H= i (a-a^\dagger)/\sqrt{2} = p\ ,
\end{equation}
is simply the momentum operator. Thus, by the spectral theorem,
\begin{eqnarray}
H_{<s} &=& P_{<s} p P_{< s} = \sum_{n,n'=0}^{s-1} \ket{n}\bra{n} p \ket{n'}\bra{n'} 
\nonumber\\
&=& \chi_{[0,E_{s-1}]}(H_{\mathrm{h.o.}}) \; p \; \chi_{[0,E_{s-1}]}(H_{\mathrm{h.o.}})\ , 
\end{eqnarray}
where 
\begin{equation}
H_{\mathrm{h.o.}} =  \frac{1}{2} (x^2 + p^2)
\end{equation}
is the harmonic oscillator quantum Hamiltonian (with  $m= \omega =1$),
whose energy is forced to be less than  $E_{s-1} = \hbar(s - 1/2)$ by the characteristic function $\chi$  [$\chi_{A}(x)=1$ if $x\in A, 0$ otherwise]. 

For large quantum numbers, we can approximate $H_{<s}$ by its classical limit, which reads
\begin{equation}
\label{Ham}
h(x,p)= p \, \chi_{[0,R]}(r)\ ,
\end{equation}
where $r=\sqrt{x^2+p^2}$ and $R=\sqrt{2 E_{s-1}}$.
This Hamiltonian describes the motion of an ultra-relativistic particle (energy proportional to momentum) confined in  phase space by a hard wall at $r=R$  (non-holonomic constraint). 

\begin{figure}
\includegraphics[width=0.5\textwidth]{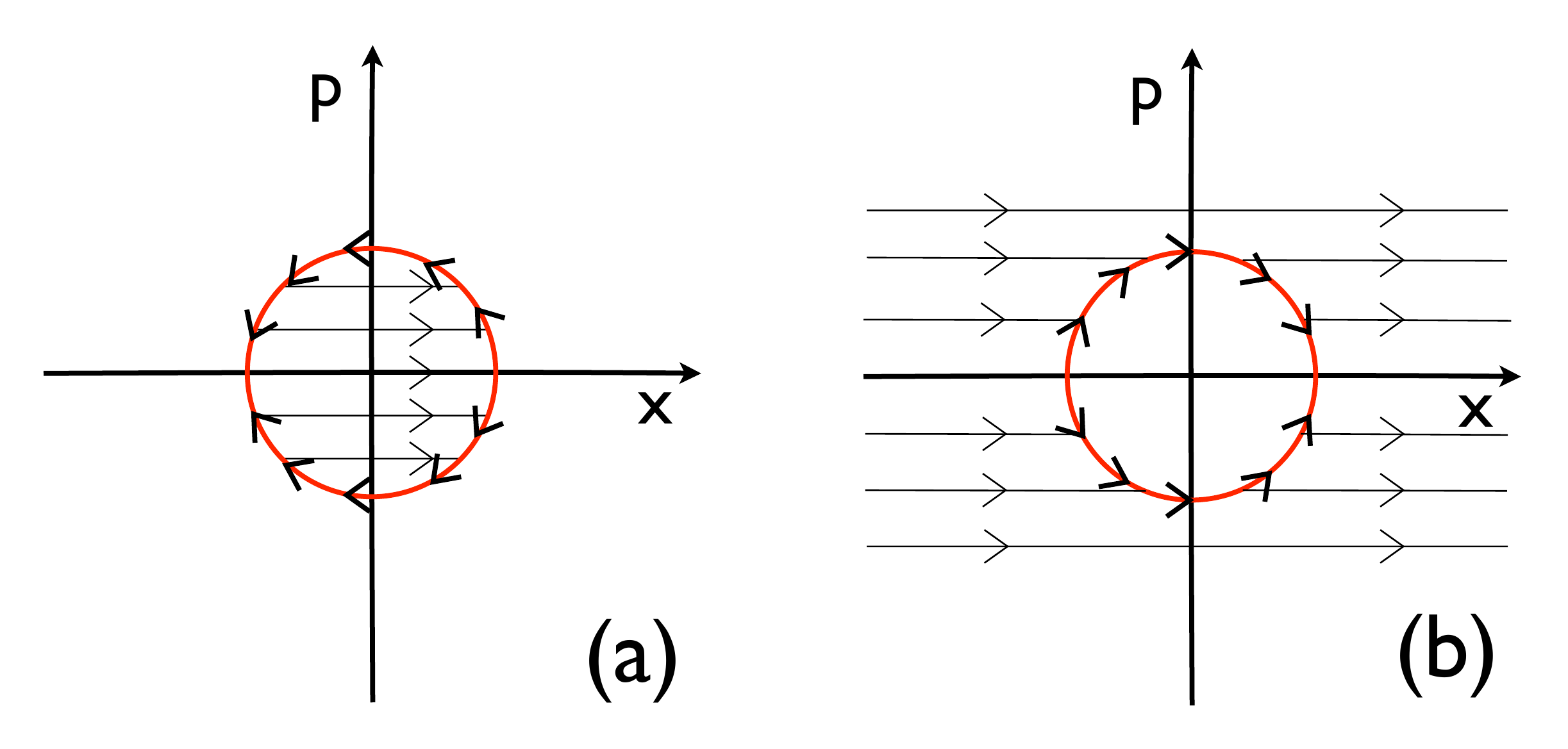}
\vspace{0cm} \caption{Vector field in classical phase space. (a) Motion inside the EC; (b) Motion outside the EC.}
\label{fluxes}
\end{figure}

The Hamilton equations of motion are:
\begin{eqnarray}
\label{Hameqs}
\dot x &=&  \chi_{[0,R]}(r) - p \delta_R(r) \frac{p}{r}\ , \\
\dot p &=&  p \delta_R(r) \frac{x}{r}\ ,
\label{Hameqs1}
\end{eqnarray}
where $\delta_R$ is the Dirac delta function at $R$.
If the particle is not on the EC, this yields the solution 
\begin{equation}
\label{solnot}
p= p_0\ , \quad x= x_0 + t\ ,
\end{equation}
$x_0$ and $p_0$ being the initial position and momentum, respectively. The particle is thus proceeding at a constant velocity along the $x$ axis. When it hits the EC, the evolution is dominated by the singular contributions in (\ref{Hameqs})-(\ref{Hameqs1}) through the vector field:
\begin{equation}
\label{solin}
X(x,p)=\frac{p}{r} \, \delta_R(r) \left(\begin{array}{c}-p \\x\end{array}\right)\ ,
\end{equation}
that yields a motion along the circle at a constant speed (infinitely large in the limit of an infinitely sharp confinement inside the EC). The particle reappears on the other side of the EC (with the same momentum) and resumes its motion along the $x$ axis at a constant velocity. These trajectories are qualitatively sketched in Fig.~\ref{fluxes}(a).

A cloud of such particles would thus evolve essentially as the field Wigner function inside the EC. This explains the `phase inversion mechanism' of Fig.\ \ref{confine}(a): the Wigner function hits the right hand side of the EC and almost instantaneously reappears on the left hand side. Of course, the transient creation of an MFSS involving a quantum superposition of two large fields with opposite phases and the appearance of an interference pattern inside the EC (Figure~\ref{confine}, frame 25) cannot be accounted for in this classical picture.

When the particle is initially \emph{outside} the EC, the evolution is generated by the Hamiltonian:
\begin{equation}
\label{Ham1}
h(x,p)= p \, (1-\chi_{[0,R]}(r))\ ,
\end{equation}
and the conclusions are identical. When the particle hits the EC, it moves very quickly to the other side  [Fig.~\ref{fluxes}(b)]. This explains the fast motion of the components of the Wigner function that come closer to the EC, in Figs.\ \ref{confine}(b) and (c).

\subsection{Long-term evolution}
\label{sec:revivals}

We now analyze the long-term evolution of the field energy when the state is initially inside the EC.
In this case, only a finite set of  Bohr frequencies is involved in the evolution, which is thus expected to 
be quasi-periodic~\cite{Exploring06}. State distortions, however, eventually accumulate and damp the oscillations of the field amplitude. 
This phenomenon was numerically investigated in~\cite{QZDcav} and will now be analyzed in greater detail.

Without loss of generality, one can consider $\alpha$ real and write:
\begin{equation}
H=\alpha (a^\dagger + a)
= \alpha \sum_{n\geq 0} \sqrt{n+1} \left(\ket{n}\bra{n+1} + \ket{n+1}\bra{n}\right)\ .
\label{eq:H2}
\end{equation}
Indeed, Hamiltonians (\ref{Hfree}) and (\ref{eq:H2}) are unitarily equivalent via $U(\varphi)= e^{i \varphi a^\dagger a}$, $\varphi$ being the phase of $\alpha$ in Eq.~(\ref{Hfree}). They have thus the same spectrum and generate the same dynamics.

Let us look first at rather small values of $s$. For $s=4$, all properties of the Zeno dynamics in $\mathcal{H}_Z$  depend on those of the matrix 
\begin{equation}
H_{<s}= 
P_{<s} H P_{<s}
=\alpha \begin{pmatrix}
0&1& 0 & 0  \\
1&0& \sqrt{2} & 0   \\
0 &\sqrt{2} &0 & \sqrt{3}   \\
0 & 0 &\sqrt{3} & 0  
\end{pmatrix}\ ,
\end{equation}
that has four non-degenerate eigenvalues
$\pm \alpha \lambda_+ =\pm  \alpha \sqrt{3+\sqrt{6}}$ and $\pm \alpha \lambda_{-} =\pm \alpha \sqrt{3-\sqrt{6}}$, so that the Zeno dynamics of a generic observable is a quasi-periodic motion on a four-dimensional torus. We get:
\begin{eqnarray}
p_0(t) &=& |\bra{0} e^{-i H_{<s} t/\hbar} \ket{0} |^2 
\nonumber\\
&=&
\frac{1}{24}\left((\lambda_+^2-1)\cos \omega  \lambda_{-} t
+ (1- \lambda_{-}^2)\cos \omega  \lambda_{+} t
\right)^2\ ,
\nonumber \\
\\
p_1(t) & = &
\frac{1}{12}\left(\lambda_+ \sin \omega  \lambda_{-} t
+ \lambda_- \sin \omega  \lambda_{+} t
\right)^2\  , \\
p_2(t) &=&
\frac{1}{12}\left(\cos \omega  \lambda_{-} t
- \cos \omega  \lambda_{+} t
\right)^2\  , \\
p_3(t) &=&
\frac{1}{12}\left(\lambda_+ \sin \omega  \lambda_{-} t
- \lambda_- \sin \omega  \lambda_{+} t
\right)^2\ ,
\end{eqnarray}
where $\omega = \alpha / \hbar$.
Thus, the population of the number states are quasi-periodic. 
We have:
\begin{equation}
\frac{\lambda_+}{\lambda_-}= 3 + \frac{1}{6 + \frac{1}{1+ \frac{1}{5 + \dots}}}\ ,
\end{equation}
so that  $\lambda_+/ \lambda_- \simeq 22/7$ up to a few per mil. Therefore,
the populations almost return to their initial value after a time
\begin{equation}
T = 22 \frac{2 \pi}{\omega  \lambda_+} 
\left(\simeq 7 \frac{2 \pi}{\omega  \lambda_-} \right) \simeq \frac{59.2}{\omega }\ .
\end{equation}

For larger values of $s$, the calculations become analytically unmanageable. However, the main conclusions remain valid and the features of the evolution qualitatively identical.

\begin{figure}
\includegraphics[width=7cm]{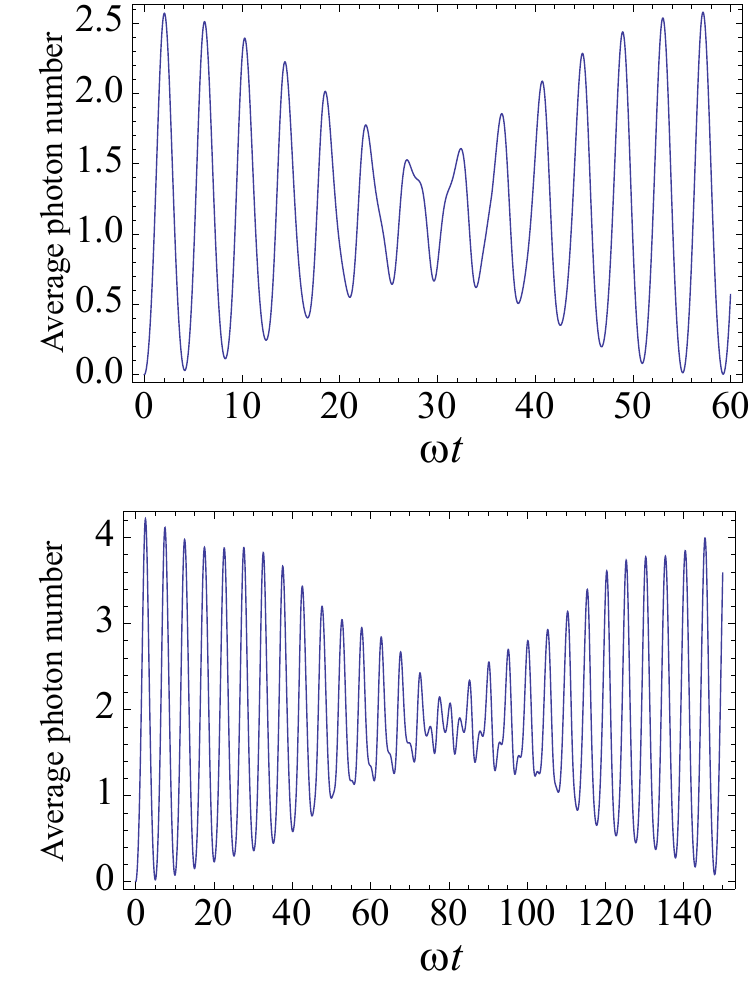}
\vspace{0cm} \caption{Number of photons as a function of $\omega  t = \alpha t / \hbar$. Upper panel: $s=4$, lower panel: $s=6$. Note in both cases the recurrence of the photon number oscillation at  $\omega  t \simeq 59.2$ and $\omega  t \simeq 150$, respectively.}
 \label{n3}
\end{figure}

The average photon numbers as a function of $\omega  t$ for $s=4$ and $s=6$ are displayed in Fig.~\ref{n3}.
We observe at long times a quantum revival~\cite{Exploring06} at $\omega  t \simeq 59.2$ and $\omega  t \simeq 150$, respectively.

\subsection{Limits of QZD and applications}
\label{sec:limits}

We have so far assumed a perfect confinement inside the EC. It can be obtained only in the limit of vanishingly small displacements $\beta$ at each step and for non-vanishing interrogation pulse Rabi angles $\phi$. In a real experiment, the preparation and interrogation take a finite time and thus the displacement per step cannot be made arbitrarily small. We have explored the corresponding limits of the QZD by extensive numerical simulations. We focus here on the typical example of a dynamics inside the EC with $s=6$ [Figure~\ref{confine}(a)]. 

The calculations have been performed using the quantum optics package for MATLAB~\cite{tan}. The field Hilbert space is truncated to the first 60 Fock states. The initial cavity state is the vacuum. Each step involves a translation by an amplitude $\beta$ (chosen real positive without loss of generality). It is followed by a Rabi interrogation pulse with an angle $\phi$ on the  $|h,6\rangle\rightarrow|+,6\rangle$ transition for an atom remaining motionless at cavity center. No atomic detection is performed and the same atom is used for all elementary steps. No other experimental imperfections (cavity relaxation, finite selectivity of the Rabi pulse, etc.) are taken into account. They will be discussed in Section~\ref{sec:simulations}. 

To assess the quality of the confinement, we compute the evolution for a number of steps $N=I[2\sqrt{6}/\beta]$ (where $I$ stands for the integer part), corresponding to the first return of the field state close to the vacuum. We compute the fidelity 
\begin{equation}
 F=\mathrm{Tr}(\varrho\varrho_p)
 \label{eq:returnfidelity}
\end{equation}
of the final field state $\varrho$ with respect to\ the reference state $\varrho_p$ obtained after the same number of steps in an ideal QZD (evolution in an Hilbert space strictly limited to the first 6 Fock states). 

\begin{figure}
 \includegraphics[width=8.5cm]{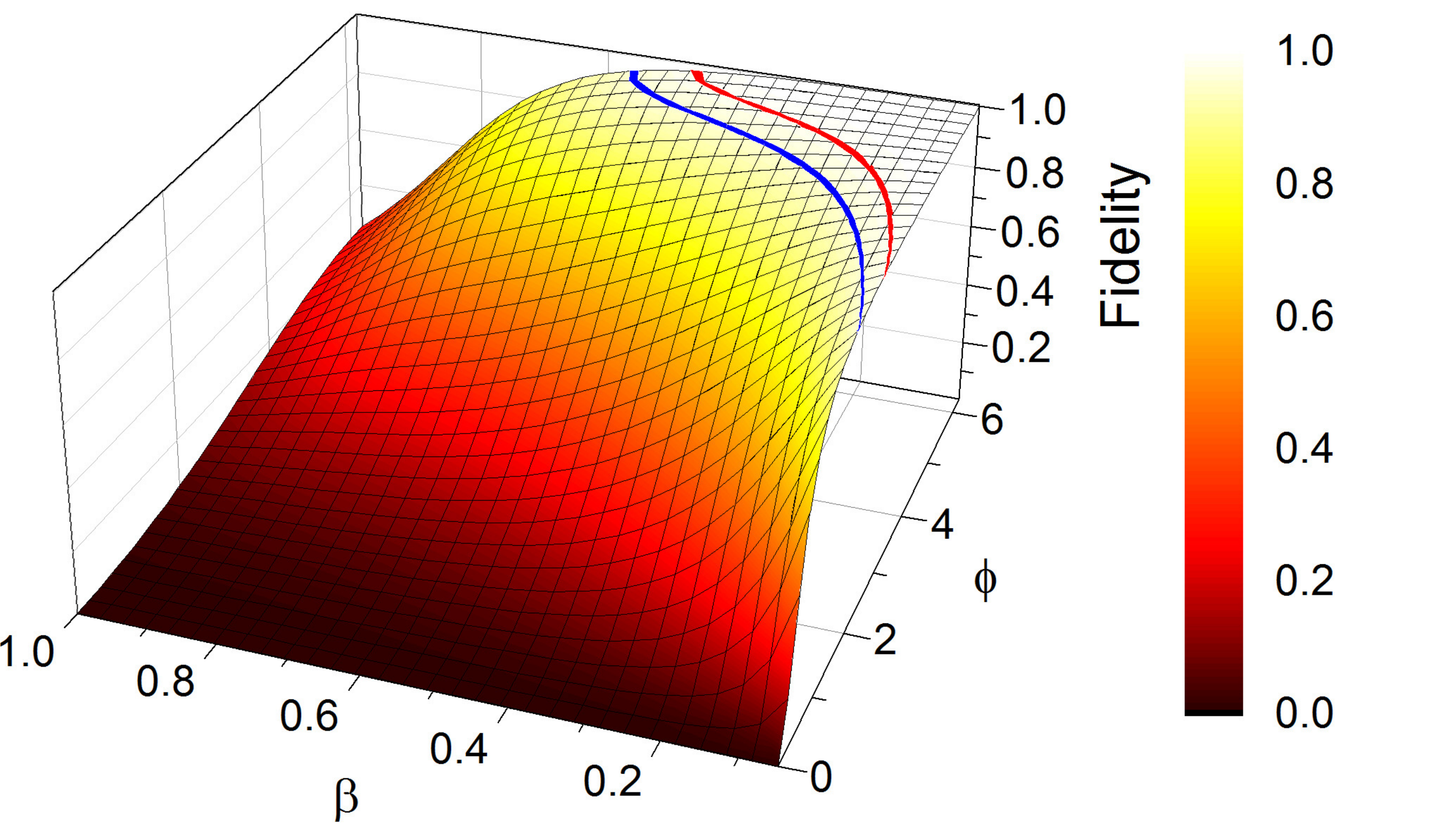}
 \caption{Fidelity of confinement for $s=6$ versus the translation per step $\beta$ and the interrogation pulse Rabi angle $\phi$ in radians. The contour lines mark the 95\% and 98\% levels}
 \label{fidelities}
\end{figure}

Figure~\ref{fidelities} presents  $F$ as a function of $\beta$ from 0.05 to 1 (0 is obviously excluded) and of $\phi$ from 0 to $2\pi$. Not surprisingly, $F$ is close to one for small $\beta$ and large $\phi$. It is nearly zero when $\phi$ approaches zero or $\beta$ one. The contours correspond to 95\% and 98\% fidelities respectively. As shown in Section~\ref{sec:simulations}, large $\phi$ values can be easily implemented in a short time interval. For a $2\pi$ interrogation pulse, we can achieve an excellent confinement fidelity (98\%) with a translation per step as large as $\beta=0.4$. The QZD is thus a quite robust mechanism. This is promising for practical applications.

When $F$ is close to 50\%, an interesting situation arises. We observe numerically that, when the collision of the moving coherent state with the EC occurs (after a number of steps around $N/2$), part of the Wigner function is transmitted through the barrier. Another part undergoes the phase inversion mechanism and is rejected to the other side of the EC. In the following steps, these two components evolve separately in the subspaces ${\cal H}_{<6}$ and ${\cal H}_{>6}$. The outer one moves further along the positive real axis and the inner one returns close to the origin. Finally we are left with two nearly coherent components centered at the origin and at an amplitude $2\sqrt{6}$.

\begin{figure}
 \includegraphics[width=8cm]{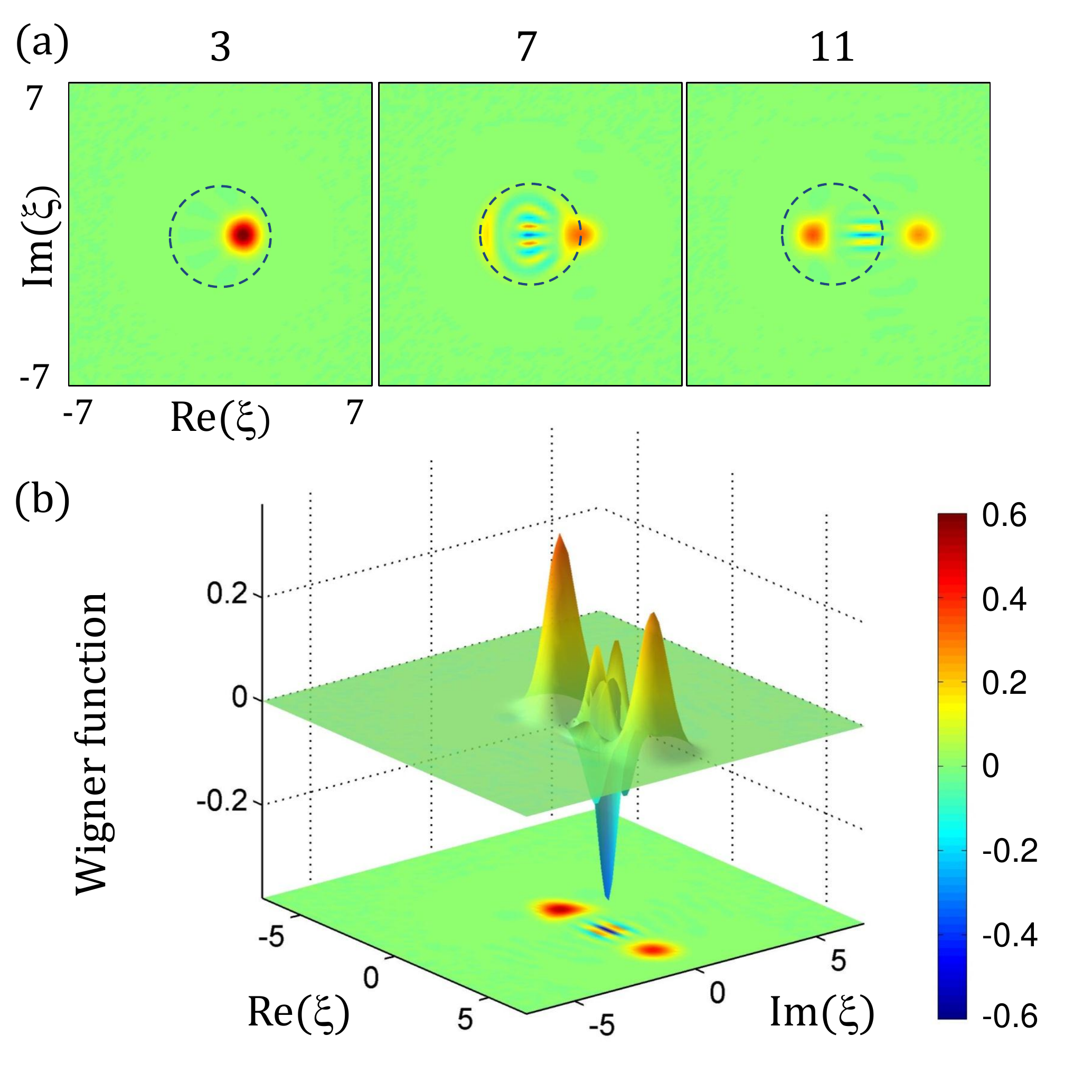}
 \caption{Generation of a cat by a semi-transparent EC ($\beta=0.345$ and $\phi=3.03$). (a) 3 snapshots of the field Wigner function $W(\xi)$. The corresponding number of steps are indicated above the frames. (b) Final  Wigner function after 14 steps. The negative parts are conspicuous.}
 \label{wignercattransparent}
\end{figure}

The final trace over the atomic state does not erase the coherence between these two components, showing that the atom is not strongly entangled with the field even though the $|s\rangle$ Fock state has been transiently populated in the process. We are thus left with a quite mesoscopic field state superposition. The evolution of the Wigner function corresponding to $\beta=0.345$ and $\phi=3.03$ is presented on figure~\ref{wignercattransparent}(a). The coherence between the two components in the final MFSS is manifest with the presence of the characteristic interference fringes [figure~\ref{wignercattransparent}(b)], even though the contrast of these fringes is not maximal.

We have systematically studied the generation of such MFSS by an imperfect QZD. For each value of $\beta$ and $\phi$, we compare the final cavity state to a superposition of two coherent states :
\begin{equation}
\ket{\text{MFSS}}=w_{<s}|\alpha_{<s}\rangle+w_{>s}e^{i\theta}|\alpha_{>s}\rangle\ ,
\end{equation} 
with real amplitudes $\alpha_{<s}$ and $\alpha_{>s}$ for the confined and transmitted parts respectively ($\alpha_{<s}$ being close to zero and $\alpha_{>s}$ close to $2\sqrt 6$). These amplitudes, the two real mixture coefficients $w_{<s}$ and $w_{>s}$, and the relative quantum phase $\theta$ are fitted to optimize the fidelity of this reference state with respect to\ the final state in the cavity. We find that the relative phase $\theta$ is always very close to $\pi$. For the conditions of figure~\ref{wignercattransparent}, the fidelity is 75\%. It is limited in particular by a residual spurious entanglement between the atom and the field.

\begin{figure}
 \includegraphics[width=8.5cm]{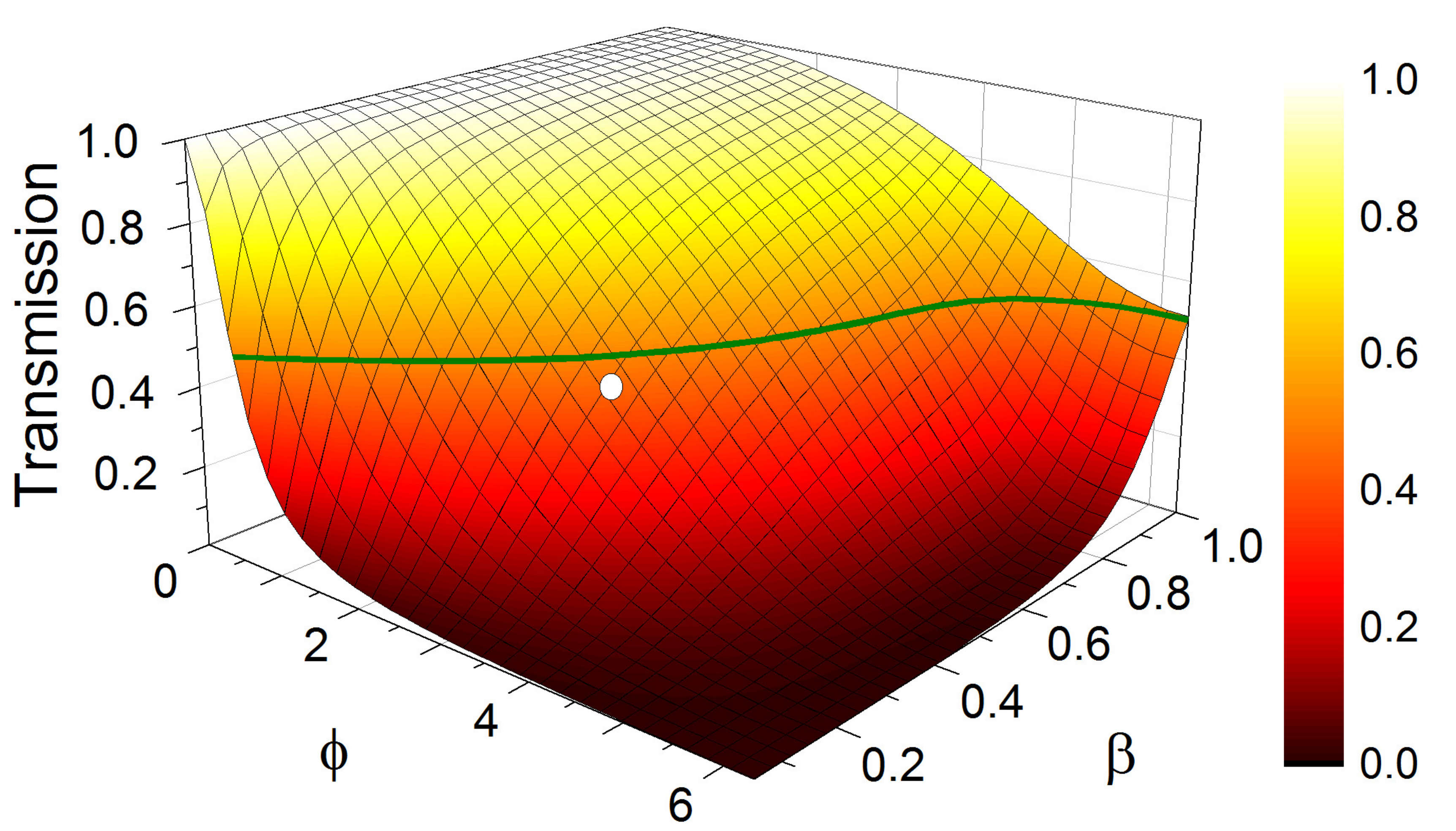}
 \caption{Transmission of the EC as a function of the QZD parameters. The solid green line follows the 50\% transmission level. The conditions used for figure~\ref{wignercattransparent} are marked by the white dot.}
 \label{transmission}
\end{figure}

 We define the EC `transparency' ${\cal T}$ as: 
 \begin{equation}
{\cal T}=w_{>s}^2/(w_{<s}^2+w_{>s}^2)\ ,
\end{equation}
weight of the transmitted component in the final MFSS. Figure~\ref{transmission} presents $\cal T$ versus the QZD parameters $\beta$ and $\phi$. The 50\% level is indicated by the green line and the conditions of figure~\ref{wignercattransparent} by the white dot (transmission 44\%). We observe that MFSS can be generated in a large range of operating parameters. Note that, for very large $\beta$ values, the state transmitted through the barrier can be notably distorted. The fidelity with respect to\ an ideal MFSS is then rather low.
 
This MFSS generation method can be straightforwardly generalized to superpositions of more than two coherent components by repeated collisions of the trapped component on a partially-transparent EC. The relative weights of these components can be adjusted by fine-tuning, during each collision, the incremental step $\beta$ and the interrogation pulse angle $\phi$. By changing the phase of $\beta$ from one collision to the next, a wide variety of multi-components superpositions can be produced.

\subsection{QZD in a translated EC}

The QZD proceeds in an EC centered at the origin of phase space. It can be straightforwardly generalized to an EC centered at an arbitrary point in phase space. Before the interrogation pulse, we perform with the help of the source $S$ a displacement of the field by a (possibly large) amplitude $-\gamma$. After the interrogation pulse, we translate back the whole phase space by the amplitude $\gamma$. Qualitatively, we block the evolution in an EC centered at the origin for a field state globally translated by the amplitude $-\gamma$. This is clearly equivalent to blocking the evolution in an EC centered at the point $\gamma$ in phase space. 

In more precise terms, the kick operator $U_K$ is changed by the two translations 
from $U_s$ into 
\begin{equation}
U_s(\gamma)=D(\gamma)U_s D(-\gamma)\ .
\end{equation}
After $p$ steps, the
global evolution operator is
\begin{equation}
U_{\mathrm{Z}}(s,\gamma,p)=[U_s(\gamma)D(\beta)]^p\ ,
\end{equation}
which can be expressed, using displacement operator commutation relations, as:
\begin{equation}
U_{\mathrm{Z}}(s,\gamma,p)=D(\gamma)U_{\mathrm{Z}}(s,0,p)D(-\gamma)\exp[2ip\Im(\beta\gamma^*)]\ .
\end{equation}
Up to a topological phase, the state after $p$ steps is
equivalently obtained by first displacing the field by $-\gamma$, then
performing $p$ QZD steps in an EC centered at origin and finally
displacing back the field by $\gamma$.

The $s=1$ case is particularly interesting in this context. The QZD blocks the unique coherent state $|\gamma\rangle$ at a fixed point in phase space, while all other parts of the phase space can be moved by the action of displacement operators. This ability to operate separately on different regions of the phase space will be instrumental in the next Section.

\section{Phase space tweezers}
\label{sec:tweezers}

We proposed in~\cite{QZDcav} to use an $s=1$ EC as phase-space tweezers. Let us assume that the initial state of the cavity field is made up of a superposition of non-overlapping coherent components, prepared for instance by using in a first stage of the experiment a semi-transparent EC. We can use an $s=1$ exclusion circle to block one of these components, with an initial amplitude $\gamma_0$. We assume here that, besides the displacements used to generate the off-center EC, there is no other source of evolution of the field. Now, we change at each step of this new QZD dynamics the center of the exclusion circle, from $\gamma_0$ to $\gamma_1$, $\ldots$, to $\gamma_N$.

Provided the difference between two successive positions of the EC, $|\gamma_{i+1}-\gamma_i|$, is always much smaller than one, the coherent component trapped in the EC will adiabatically follow the motion of its center. The coherent state amplitude will thus be changed from $\gamma_0$ to $\gamma_N$, while all other components of the initial state remain unchanged.

The movable EC operates in phase space as the optical tweezers which are now routinely used to move microscopic objects. In analogy, we coined the term `phase space tweezers' for this operation~\cite{QZDcav}.

Obviously, ideal operation of the phase space tweezers requires an infinitely small increment of the EC position at each step, hardly compatible with a practical implementation. We have thus studied the quality of the tweezers operation with respect to the interrogation pulse characteristics and to the `velocity' of the EC motion.  

We compute the final field state for a tweezers action, taking initially the vacuum state and pulling it away. As in Section~\ref{sec:limits}, all sources of experimental imperfections are neglected in this calculation. The exclusion circle center moves in $N$ steps from zero to a real amplitude $2\sqrt 6$ (i.e. 24 photon field, this amplitude being chosen rather arbitrarily to coincide with the total field displacement used in Section~\ref{sec:limits}). We compute the final state fidelity with respect to\ a coherent state with an optimized amplitude. We observe in fact that, for small $N$ values i.e. large EC displacements per step, the state wiggles slightly inside the EC during translation. The final amplitude might thus not be exactly equal to $2\sqrt 6$. The effect is quite small, the maximum amplitude difference being less that 0.5 for all data presented here

\begin{figure}
 \includegraphics[width=8.5cm]{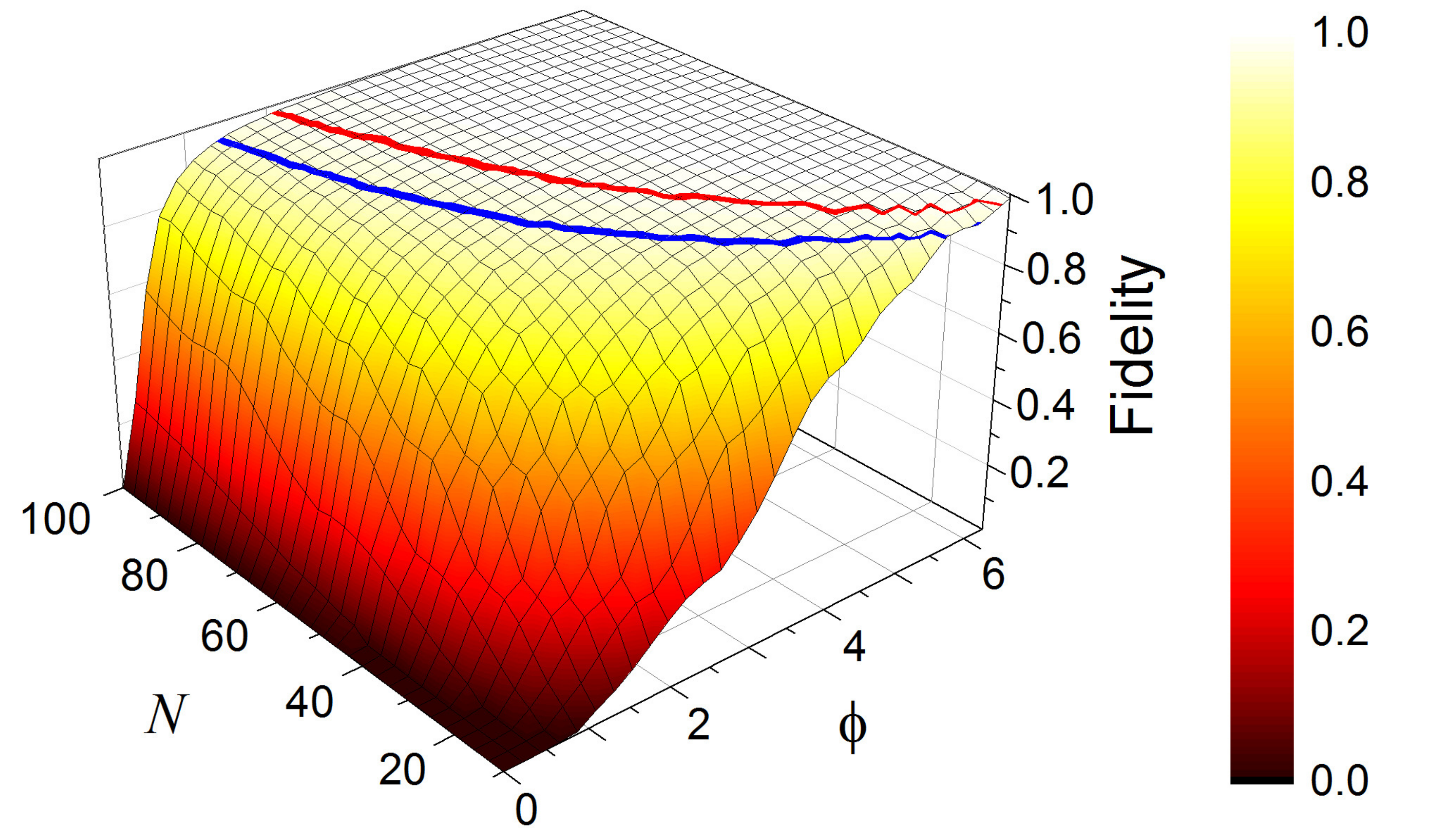}
 \caption{Tweezers operation fidelity as a function of $N$ and $\phi$. The fidelity is computed with respect to an optimized coherent state. The two solid lines follow the 95\% and 98\% levels.}
 \label{fidelitiestweezer}
\end{figure}

Figure~\ref{fidelitiestweezer} presents the calculated final fidelity as a function of the number of steps $N$ (from 10 to 31) and of the interrogation pulse Rabi angle $\phi$ (from zero to $2\pi$). The 95\% and 98\% levels are indicated by the blue and red contour lines. For the $2\pi$ interrogation pulse, the fidelity is extremely large, more than $99\%$ for $N=10$, i.e. a motion of nearly 0.5 per step. Once again, the QZD operation is very robust and tweezers can be used to move quite rapidly a coherent component through the phase plane. 

\begin{figure}
\includegraphics[width=8.5cm]{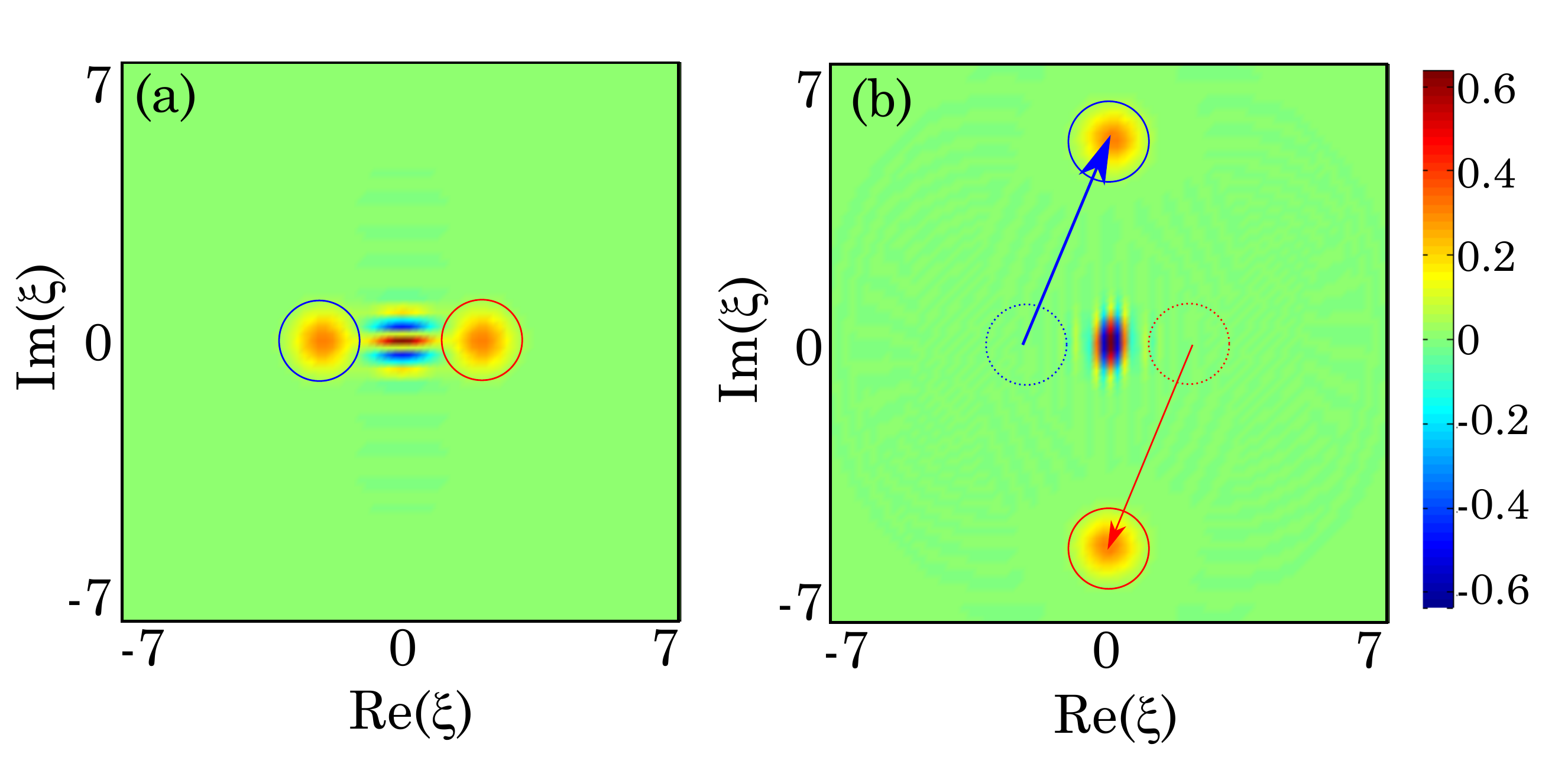}
\caption{(a)-(b) Initial and final Wigner functions $W(\xi)$ for a
phase space tweezers operation. The first steps ECs are depicted as
solid lines in (a) and dotted lines in (b), the final ECs as solid
lines in (b). The arrows in (b) indicate the two EC centers
trajectories. From~\cite{QZDcav}. }
 \label{tweezer}
\end{figure}

Obviously the fidelity decreases quite rapidly with the interrogation angle $\phi$. For $\phi=\pi$, $N=60$ steps are required to achieve a 99\% fidelity. When the EC step is too large, or the interrogation angle to low, the EC is slightly transparent at each step and leaks a little bit of the trapped state. The final state is stretched along the path of the EC, with no remarkable features. 

Phase space tweezers can be used to increase at will the distance between two components of an MFSS as illustrated in figure~\ref{tweezer}. The initial cavity state is $|\alpha\rangle+|-\alpha\rangle$ with $\alpha=2$.
It is turned in 100 steps (50 for the motion of each component) into $|\alpha'\rangle+|-\alpha'\rangle$ with $\alpha'=5i$. The final
fidelity is $98.8$\% with respect to the expected cat. A wide variety of operations on non-classical fields can be envisioned with this concept.

The tweezers operation allows us to tailor at will a pre-existing superposition of coherent states. It can be slightly modified to allow for the generation of this superposition from the vacuum, as shown in the next Section.

\section{State synthesis}
\label{sec:synthesis}

We present in this Section a QZD-based method for the generation of a nearly arbitrary superposition of coherent components, starting from the vacuum state. We proceed with an EC motion driving the vacuum to the first required coherent component. However, we perform a quantum superposition of the tweezers operation and no operation at all by casting the interrogation atom in a superposition of $h$ with an inactive state $i$  (for instance the circular state with principal quantum number 52, lying above $e$), which does not take part in the QZD process. The atom gets in the process entangled with a field involving a superposition of the vacuum with a moving coherent component. This process is then repeated for all the coherent components in the final state.

Let us write the target state $|\Psi_t \rangle$ as: 
\begin{equation}
 |\Psi_t \rangle=\sum_{j=1}^m c_j|\gamma_j\rangle\ ,
 \label{eq:psit}
\end{equation}
superposition of $m$ coherent states. We assume that all the $\gamma_j$s with $j\not = m$ have a negligible mutual overlap as well as with the vacuum state. Up to an irrelevant global phase, we can also assume $c_m$ real.

We first create the component $|\gamma_1\rangle$ out of the initial cavity vacuum state $|0\rangle$. The atom, initially in $h$, is at rest at cavity center. We send on the atom, with the help of a microwave source $S_2$, a narrow-band (soft) microwave pulse resonant with the $|h,0\rangle\rightarrow|g,0\rangle$ transition, tuned to produce the state superposition $a_1|g,0\rangle+b_1|h,0\rangle$ (the $a_1$ and $b_1$ coefficients will be determined later).

A tweezers operation performed with the atom initially in $g$ and an empty cavity leads to a partially transparent EC (the initial state has a component on $|-,0\rangle$, which is not addressed by the interrogation pulses) and to a spreading in phase space. 
We must avoid this effect. Before performing the tweezers action, we thus shelve level $g$ in the fourth level $i$. We use for this purpose a millimeter-wave source $S_3$ tuned to resonance with the two-photon transition between $g$ and $i$ at 2$\times49.6$ GHz. The strong coupling of Rydberg atoms to millimeter-wave sources makes it possible to achieve a $\pi$ pulse on this transition in a short time interval. Such a short (hard) pulse does not resolve the dressed level structures and performs the transition whatever the photon number in the cavity. Finally, we reach the quantum state superposition $a_1|i,0\rangle+b_1|h,0\rangle$. 

We then perform the tweezers action itself, using the interrogation source $S'$ tuned for $s=1$ and the translation source $S$. The tweezers is active only if the atom is initially in state $h$. The EC center evolving from 0 to $\gamma_1$, we are finally left with the entangled atom-cavity state  $a_1|i,0\rangle+b_1|h,\gamma_1\rangle$. We do not take into account here any topological phase that could affect the $|h,\gamma_1\rangle$ part of the state if the trajectory through phase space was not a straight line. This phase could easily be taken into account with minor modifications of the algebraic expressions. A final hard $-\pi$ pulse on the $i\rightarrow g$ transition driven by $S_3$ leads us to $a_1|g,0\rangle+b_1|h,\gamma_1\rangle$.

Since $\gamma_1$ is notably different from zero, a soft pulse on the $|h,0\rangle\rightarrow|g,0\rangle$ transition driven by $S_2$ addresses only the part of the atom-cavity state involving the vacuum. We tune this pulse to produce the state superposition $a_1(a_2|g,0\rangle+b_2|h,0\rangle)+b_1|h,\gamma_1\rangle$. We then shelve $g$ in $i$ with a hard pulse driven by $S_3$ and perform a tweezers operation leading from the vacuum to the amplitude $\gamma_2$. We should take care that the EC never comes close to the $\gamma_1$ component, which should be left unchanged. There is of course ample space in the phase plane to plan a convenient trajectory. Finally, we unshelve level $i$, leading to the state $a_1a_2|g,0\rangle+a_1b_2|h,\gamma_2\rangle +b_1|h,\gamma_1\rangle$, involving a superposition of three coherent components, two of them ($\gamma_1$ and $\gamma_2$) being disentangled from the atomic state.

Since again $\gamma_2$ is notably different from zero, we can selectively address with $S_2$ the $g$ part of the state to split it in a coherent superposition. Iterating the process $m$ times, we prepare finally the state:
\begin{eqnarray}
 &&a_1a_2\cdots a_{m-1}|g,0\rangle+a_1a_2a_3\cdots a_{m-2} b_{m-1}|h,\gamma_{m-1}\rangle+
 \nonumber\\&&\ldots +a_1b_2|h,\gamma_2\rangle +b_1|h,\gamma_1\rangle\ .
\end{eqnarray}
A final $\pi$ pulse produced by $S_2$ on $|g,0\rangle\rightarrow|h,0\rangle$ casts the atom in $h$ with certainty and a final tweezers operation from $0$ to $\gamma_m$ leaves the atom in $|h\rangle$ and the cavity in the state:
\begin{eqnarray}
 &&a_1a_2\cdots a_{m-1}|\gamma_m\rangle+a_1a_2a_3\cdots a_{m-2} b_{m-1}|\gamma_{m-1}\rangle
 \nonumber\\&&\ldots+ a_1a_2b_3|\gamma_3\rangle+a_1b_2|\gamma_2\rangle +b_1|\gamma_1\rangle\ .
\end{eqnarray}

We must now determine the intermediate coefficients $a_i$ and $b_i$ so that the final state is $|\Psi_t\rangle$ [Eq.~(\ref{eq:psit})]. The simplest choice is obtained by setting $b_1=c_1$ and $a_1=\sqrt{1-|b_1|^2}$. This determines the value of $b_2$ and hence (within an irrelevant phase that we take to be zero), that of $a_2$. We then get $b_3$ and $a_3$ and stepwise all the required coefficients.

\begin{figure}
\includegraphics[width=7.5cm]{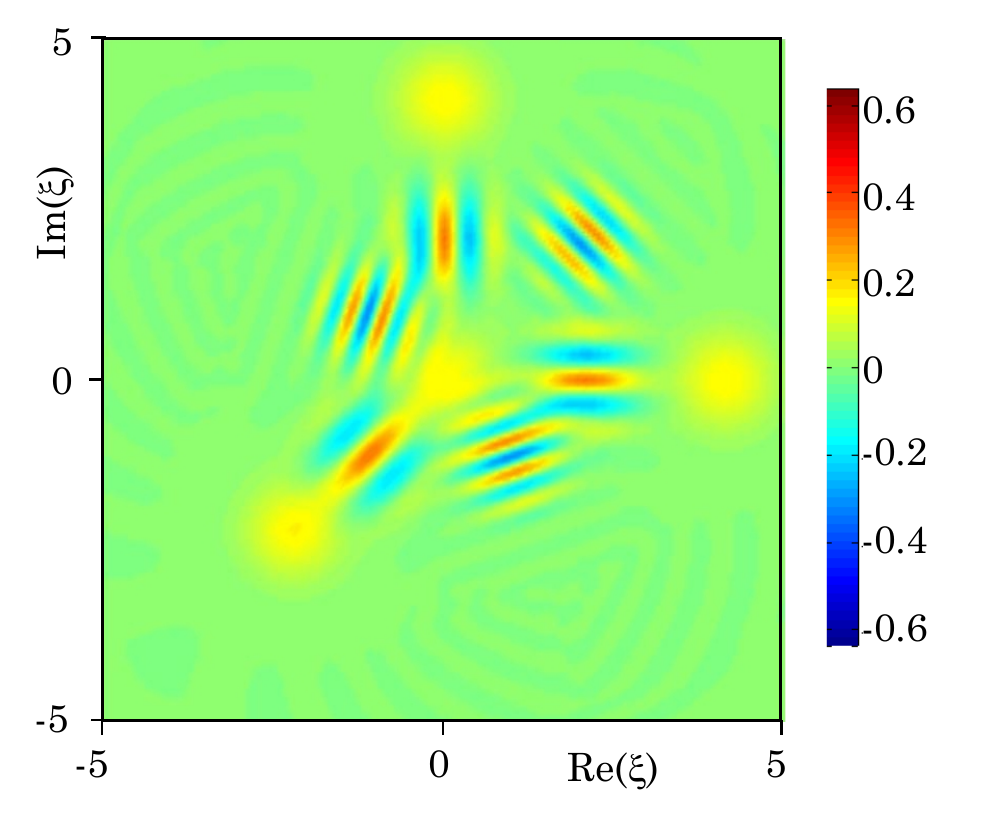}
\caption{Final state Wigner function $W(\xi)$ after the synthesis of a complex state superposition. See text for the conditions.}
\label{fourlegged}
\end{figure}

We have numerically simulated the procedure for the creation of a four-component MFSS: 
\begin{equation}
 \frac{1}{2}\left(|4\rangle+|4i\rangle+|3e^{i5\pi/4}\rangle+|0\rangle\right)\ .
\end{equation}
All tweezers actions are performed with a 0.1 amplitude increment and a $\phi=2\pi$ interrogation pulse. The Wigner function of the resulting state is plotted in figure~\ref{fourlegged}. The fidelity with respect to\ the target state is 99\%.

This method opens many perspectives for the generation of complex MFSS. The only restriction is that the final components should not overlap with each other and with the vacuum (if this is not the case, it is not possible to manipulate one independently of the others with the tweezers). There is nevertheless a wide range of state superpositions that can be directly reached with this method.

\section{Simulations of a realistic experiment}
\label{sec:simulations}

We have up to now discussed the QZD in an ideal setting, assuming no atomic motion, no cavity relaxation and, more importantly, a perfect selectivity of the interrogation pulse. We now proceed to include a realistic description of these imperfections and to assess the quality of the QZD in this context. 

First, atomic motion through the mode, at a low velocity, is not a real problem. Provided the initial position of the atom (determined by the excitation lasers) and the atomic velocity are  well known, the position of the atom is precisely known at any time during the sequence. The slow variation of the atom-cavity coupling can then be taken into account. We can for instance tune the interrogation pulse source to remain resonant on the selected dressed atom transition.

We thus address in more details the two other issues. The main one is the interrogation pulse selectivity, discussed in the next Subsection. The final Subsection is devoted to realistic simulations of a few key QZD experiments including cavity relaxation.

\subsection{Interrogation pulse optimization}

The interrogation pulse should address selectively the $|h,s\rangle\rightarrow|+,s\rangle$ transition. For a motion inside the EC, or for a tweezers operation, this pulse should not affect any transition corresponding to a photon number $n<s$, and more particularly the transition between $|h,s-1\rangle$ and $|+,s-1\rangle$, which is closest to resonance with the interrogation pulse. The frequency difference between the addressed and the spurious transitions is only $\delta_n=(\Omega/2)|\sqrt{s}-\sqrt{n}|$. In particular, $\delta_{s-1}$ decreases with increasing $s$. 

For the sake of definiteness, we shall only consider two practical cases in the following, that of a tweezers operation ($s=1$) and that of a motion inside the EC with $s=6$. In the latter case, the interrogation pulse should resolve a frequency splitting $\delta_5/2\pi=5.3$~kHz only. Relying on a very long pulse duration to achieve this resolution leads to unrealistically long experimental sequences in view of the finite cavity lifetime. A careful tailoring of the interrogation pulse is the only realistic solution.

\subsubsection{Square pulses}

The simplest procedure is to set $S'$ to produce a square pulse with a duration $t_p$, resonant with the addressed transition and performing a Rabi rotation by an angle $n_p\pi$. In most cases, $n_p$ is set to two, but smaller values can be used ($n_p=1$ is appropriate to implement a semi-transparent EC -- note that $n_p$ need not be an integer). To minimize unwanted transitions, we can chose the pulse duration so that the same pulse produces a $p_p\pi$ Rabi pulse on the non-resonant nearby transition ($n=s-1$), where $p_p$ is an even integer (obviously larger than $n_p$). This condition sets a zero in the spectrum of the pulse at the  precise frequency of the spurious transition.

A simple algebra on Rabi rotations leads to a pulse duration $t_p=(\pi/\delta_{s-1})\sqrt{p_p^2-n_p^2}$. For $s=1$ and $n_p=2$, we get $t_p=69\ \mu$s for $p_p=4$ and $t_p=113\ \mu$s for $p_p=6$. In the more demanding $s=6$ case, the pulse durations with the same settings are 324 and 530 $\mu$s respectively. All these durations are still much shorter than the atom-cavity interaction time scale.

Numerical estimations of the influence of this square pulse on the complete dressed states structure confirm that the transfer rates on the non-resonant transitions are small, in the \% range at most for the shortest pulses. However, we observe that the relatively strong pulse produces an appreciable phase shift of the states $|h,n\rangle$ with $n<s$, due to the accumulated light shift effect. This is not a too severe problem for the tweezers operation, since this amounts finally to a global, predictable, phase shift on the displaced component. This phase shift could in principle be compensated for or taken into account in the state synthesis method.

The influence of these light-shifts is, however, much more obnoxious for the QZD in an EC. Setting $s=6$ and $p_p=4$, we get a $-0.86$~rad shift for $|h,5\rangle$, $-0.47$~rad for $|h,4\rangle$. This phase shift is inversely proportional to $\delta_n$. It is not a linear function of $n$ and cannot be absorbed, as an index of refraction effect, in a mere redefinition of the cavity frequency. We have checked by numerical simulations that this phase shift destroys most of the QZD features. The EC remains an impenetrable barrier, but the state inside it is completely distorted, even far before it reaches the EC for the first time.

Note that pulse shape optimization can reduce the spurious transfer rates but does not solve the light shifts problems. A more sophisticated pulse sequence is mandatory.

\subsubsection{Optimized composite pulses}

We propose thus to use a composite pulse sequence that leads to a nearly perfect cancellation of the light shifts. We discuss it in the important case of a $2\pi$ interrogation pulse. The sequence is made up of three pulses:

\begin{itemize}
\item a $\pi$ square pulse on $|h,s\rangle\rightarrow|+,s\rangle$, carefully optimized as in the previous Subsection ($n_p=1$, $p_p=2$ or 4).

\item a fast phase shift of the atomic levels alone, changing $|g\rangle$ into $-|g\rangle$ and amounting to exchanging the $|+,n\rangle$ and $|-,n\rangle$ dressed states for all photon numbers.

\item an optimized $\pi$ pulse on $|-,s\rangle\rightarrow|h,s\rangle$, similar to the first one within an adjustable phase $\varphi$.
\end{itemize}

In principle, the first pulse transfers all the population from $|h,s\rangle$ to $|+,s\rangle$. The central phase shift transforms $|+,s\rangle$ into $|-,s\rangle$. The final pulse transfers back the state into $|h,s\rangle$, with a phase shift that can be adjusted by tuning the phase $\varphi$. For all other levels ($n\not = s$), there is almost no transfer out of $|h,n\rangle$ by the initial and final optimized pulses and the central operation has thus no effect. In this composite sequence, the two microwave pulses applied on the atom-cavity system have opposite detunings with respect to the spurious $|h,n\rangle\rightarrow|+,n\rangle$ and $|-,n\rangle\rightarrow|h,n\rangle$ transitions ($n<s$). One can thus expect that the phase shifts due to the second pulse exactly compensate those produced by the first.

The central phase shift operation could be performed by a hard non-resonant pulse coupling $g$ to another level ($i$ for instance). The accumulated light shift can be tuned for an exact $\pi$ phase shift of $g$, independent upon the photon number in the cavity. Levels $h$ and $e$, farther away from resonance with this dressing pulse, are not affected. A simpler solution is to use the differential  Stark shift on the three levels $e$, $g$ and $h$ as in~\cite{Meunier05}. A short pulse of electric field applied across the cavity mirrors  produces three different photon-number-independent phase shifts, $\varphi_e$, $\varphi_g$ and $\varphi_h$ on these levels. Setting $\varphi_e-\varphi_g=\pi$, we are left within a global phase with the transformations $|e\rangle\rightarrow|e\rangle$, $|g\rangle\rightarrow-|g\rangle$ and $|h\rangle\rightarrow e^{i\Phi}|h\rangle$. For most QZD operations, the phase $\Phi$ acting on $h$ is an irrelevant global phase factor. For the state synthesis, it is a well-known quantity that can be taken into account in the  state preparation sequence.

We have evaluated numerically the effect of this composite pulse. For $s=1$, $n_p=1$, $p_p=4$, the total duration of the sequence is 155~$\mu$s. With $\varphi=2.75$~rad, we perform the selective transformation $|h,1\rangle\rightarrow-|h,1\rangle$. The residual phase shift on $|h,0\rangle$ is only $-3.\,10^{-5}$~rad.  In the more demanding $s=6$ case, we use $n_p=1$, $p_p=2$ for a total duration of 325~$\mu$s. The residual transfer rates for $n<s$ are below 1.5\% and the residual phase shifts lower than $10^{-4}$~rad. For the few tens of pulses in a typical QZD sequence, these imperfections have a negligible influence. 

The principle of this composite pulse can be extended to other $\phi$ values and, in particular, to $\phi=\pi$, useful for the realization of semi-transparent ECs. This pulse is made up of an $n_p=1/2$ pulse on $|h,s\rangle\rightarrow|+,s\rangle$, followed by the atomic phase inversion and a $n_p=1/2$ pulse on $|-,s\rangle\rightarrow|h,s\rangle$ with a phase $\varphi$. In the ideal case, this pulse combination results in the transformations $|h,s\rangle\rightarrow|-,s\rangle$; $|-,s\rangle\rightarrow|+,s\rangle$ and $|+,s\rangle \rightarrow|h,s\rangle$ (three applications of the transformation are necessary to return to $|h,s\rangle$). Since the level $|h,s\rangle$ is nearly never populated in a successful QZD, this composite pulse is basically equivalent to a standard $\pi$ pulse.

The composite pulse architecture achieves the required selectivity in a relatively short interrogation time. We use now these optimized pulses for the simulation of a few key QZD experiments.

\subsection{Simulation of key experiments}

The simulations include a realistic description of the composite interrogation pulse and of cavity relaxation. We use the best available cavity damping time $T_c=130$~ms~\cite{cavitytechnique}. We should of course check that the total duration of the sequence remains in the ms time range, much shorter than the atomic free fall through the cavity mode.

The periodic motion of a coherent component inside the EC is barely affected by cavity relaxation. Setting $s=6$, using a composite $2\pi$ interrogation pulse with a translation per step $\beta=0.4$, we get a fidelity after 12 steps (corresponding to the return near the vacuum state) of 90\% instead of 92\% in the ideal case treated in Section~\ref{sec:confined}. The total sequence duration is 3.9~ms. In fact, the field propagates most of the time inside the EC as a coherent state, nearly impervious to relaxation. It is only during the phase inversion, for a few steps, that a MFSS prone to decoherence is generated.

We have also examined the creation of a MFSS by transmission trough a semi-transparent EC. With a composite $\pi$ interrogation pulse, $s=6$ and $\beta=0.33$, we obtain in 5.4~ms a fidelity with respect to\ an ideal state of 79\%. This is promising to study the decoherence of this large MFSS (the square of the distance in phase space between the two components, setting the decoherence time scale~\cite{Exploring06}, is 24 photons).

\begin{figure}
 \includegraphics[width=7cm]{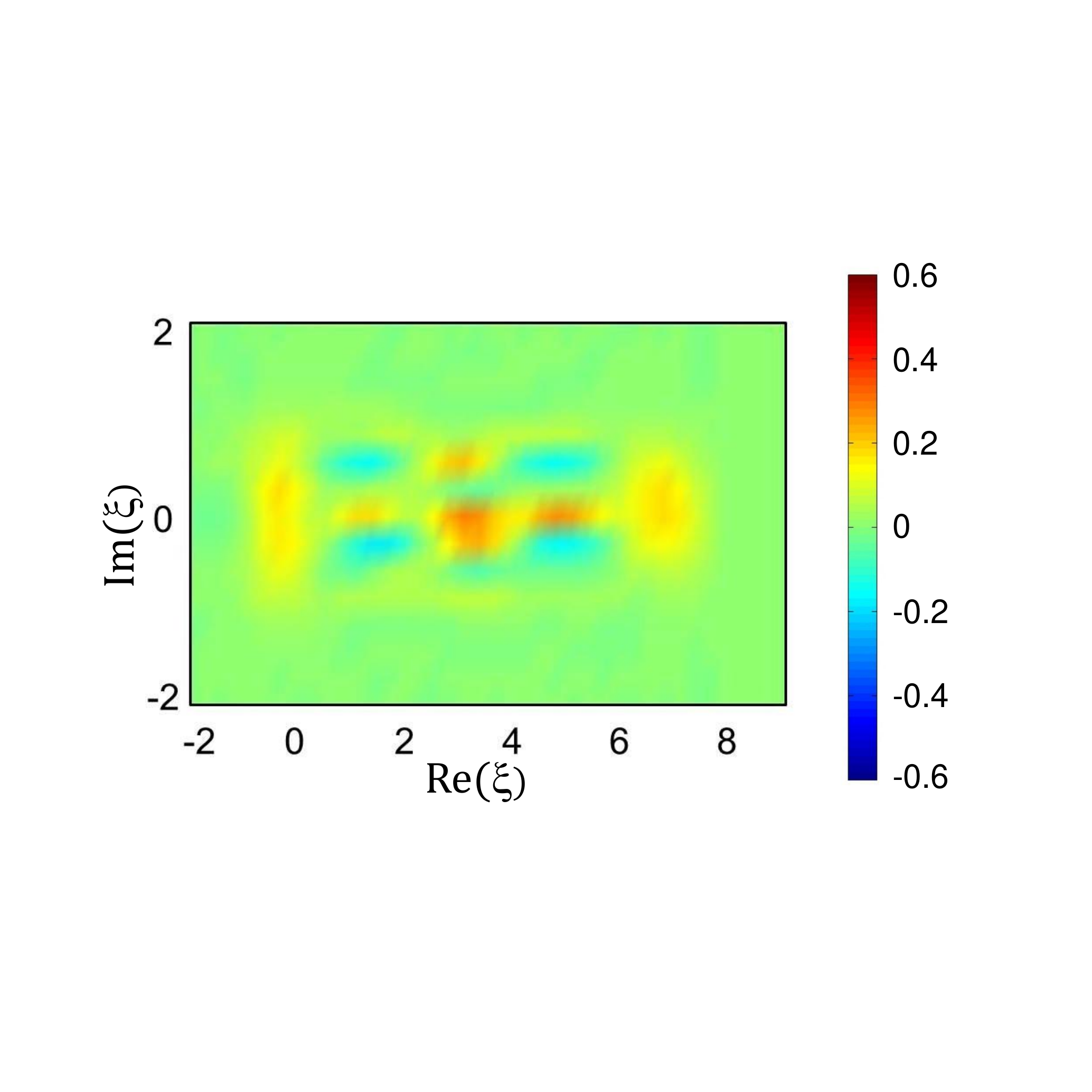}
 \caption{Final state Wigner function $W(\xi)$ for the generation of a three-component cat. See text for the conditions.}
 \label{threeleggedcat}
\end{figure}

As a more striking example, we have simulated the generation of a three-component MFSS by two collisions with a semi-transparent EC. We set $s=3$ and use a composite $\pi$ pulse. In order to get a superposition with three equal weights, we select an EC transparency of 1/3 for the first collision ($\beta=0.34$) and 1/2 for the second, setting $\beta=0.45$ after the first phase inversion. The sequence duration is 4.4~ms. The Wigner function of the final state is plotted in figure~\ref{threeleggedcat}. The fidelity with respect to an equal weight superposition of coherent components centered at $-0.3$, 3.2 and 6.7 is 69\%.

Finally, we have simulated the state synthesis presented in
Section~\ref{sec:synthesis}, leading to a MFSS of four coherent
components. The pulses addressing the $|h,0\rangle\rightarrow|g,0\rangle$
transition can spuriously affect the $|h,n\rangle$ levels with $n\ge 1$.
The frequency separation between the addressed and spurious transitions
is quite large in this case ($\Omega/2$ for $n=1$). We use thus simply
an optimized square pulse, performing the required level mixing on the
addressed transition and a $4\pi$ pulse on the closest spurious
transition. The maximum pulse duration involved in the sequence is 80$\ \mu$s. For the interrogation of the dressed states, we use the
optimized composite pulses. The tweezers operations are performed with a
$\beta=0.6$ translation per step. The total duration of the full synthesis sequence is
thus 2.9~ms.

\begin{figure}
\includegraphics[width=7cm]{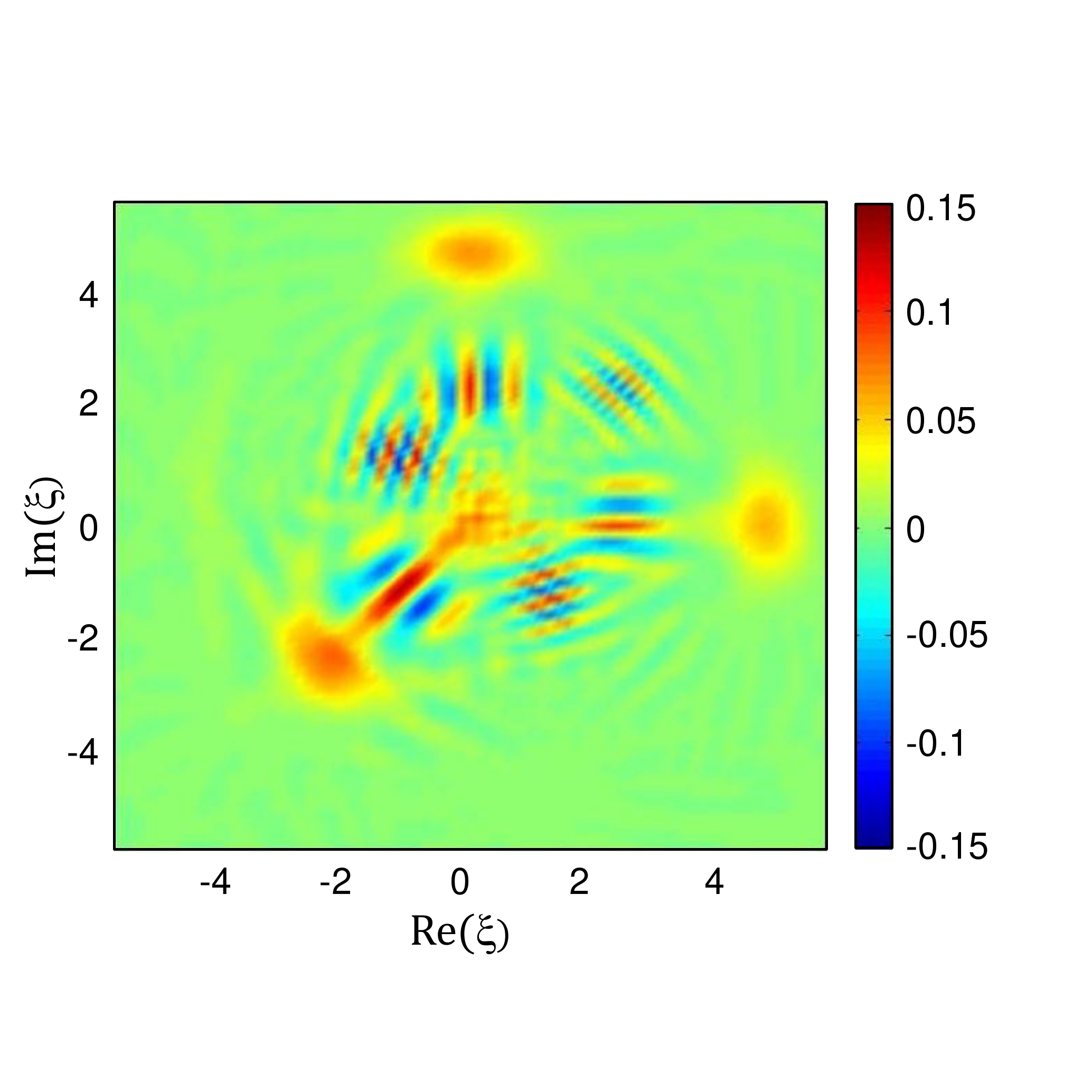}
\caption{Final state Wigner function $W(\xi)$ for the realistic state
synthesis. See text for the conditions.}
\label{realsynthesis}
\end{figure}

Figure~\ref{realsynthesis} presents the Wigner function of the generated
MFSS. It is visually very similar to the ideal MFSS Wigner function
presented in Fig.~\ref{fourlegged} (note the different color scales).
The fidelity with respect to\ the target state is 59\%. In fact, due to the fast
twezers operations, the final amplitudes of the coherent
components differ by up to 0.5 from the target ones. By optimizing these
amplitudes in the reference state, we get a more faithful fidelity of
71.5\%. This value shows that a complex state synthesis operation
is within reach of the planned experimental setup.

\section{Conclusion and perspectives}
\label{sec:concl}

We have analyzed the Quantum Zeno dynamics taking place when the photon field in a high-finesse cavity undergoes frequent interactions with atoms, that probe its state, yielding photon-number selective measurements or unitary kicks. A coherent classical source induces an evolution of the field, which remains confined in a multidimensional eigenspace of the measurement or kick. The quantum coherence of the evolution under the action of the source is preserved, the generator of the dynamics being the Zeno Hamiltonian, projection of the complete source-induced Hamiltonian onto the eigenspaces of the measurement or kick operators.

The QZD evolution can be highly non-trivial. We have discussed in particular the generation of interesting non-classical states, including MFSS. We have also analyzed state manipulation techniques by means of phase space tweezers, as well as promising perspectives towards quantum state synthesis. These ideas pave the way towards more general phase space tailoring and `molding' of quantum states, which will be of a great interest for the exploration of the quantum-to-classical transition and for the study of non-trivial decoherence mechanisms.

We have focused in this paper on the QZD induced by a classical resonant source acting on the cavity. Other evolution Hamiltonians could be envisioned, such as a micromaser evolution~\cite{micromaser} produced by fast resonant atoms crossing the cavity between the interrogation pulses performed on the atom at rest in the mode. The principle of the method could also be translated in the language of any spin and spring system. In particular, QZD could be implemented using this method in ion traps~\cite{ion} or in circuit QED~\cite{circuitQED} with superconducting artificial atoms.

In conclusion, a state-of-the-art experiment appears to be feasible in microwave cavity QED. It would be the first experimental demonstration of the quantum Zeno dynamics.

\begin{acknowledgments}
We acknowledge support by the EU and ERC (AQUTE and DECLIC projects) 
and by the ANR (QUSCO-INCA).
\end{acknowledgments}

\end{document}